\documentclass[12pt]{article}
\usepackage{amsmath, amsthm,comment}
\usepackage{amsfonts,color}
\usepackage{amssymb,graphics,psfrag}
\usepackage{array,epsfig,stmaryrd,graphicx}
\usepackage{hyperref}
\usepackage{datetime}
\usepackage[export]{adjustbox}

\pdfoutput=1
\usepackage[utf8]{inputenc}

\newcommand{\sltr}{\mathrm{SL}(2,\mathbb{R})}

\newenvironment{eqaed}
    {
    \begin{equation}
        \begin{aligned}
    }
    {
        \end{aligned}
    \end{equation}\ignorespacesafterend
    }

\usepackage{amsmath,amsthm,amssymb}
\usepackage[dvipsnames]{xcolor}
\usepackage{physics}
\usepackage{tensor}
\usepackage{graphicx}
\usepackage{float}
\usepackage{xcolor, colortbl}
\usepackage{hyperref}
\usepackage{authblk}
\usepackage{mathtools}
\usepackage{geometry}
\definecolor{light-gray}{gray}{0.9} 

\usepackage{setspace}
\usepackage{xparse}
\usepackage{esvect} 
\usepackage{slashed} 
\usepackage{dsfont} 
\usepackage[dvipsnames]{xcolor} 
\usepackage{empheq}
\usepackage{makecell}
\usepackage{subcaption}
\usepackage{float}
\usepackage{hyperref}
\hypersetup{
    colorlinks=true,
    linkcolor=blue,
    filecolor=magenta,      
    urlcolor=cyan,
}
\usepackage{import}
\usepackage{makeidx}

\newcommand{\beq}{\begin{equation}}
\newcommand{\eeq}{\end{equation}}
\newcommand{\beqa}{\begin{eqnarray}}
\newcommand{\eeqa}{\end{eqnarray}}
\newcommand{\beqar}{\begin{eqnarray*}}
\newcommand{\eeqar}{\end{eqnarray*}}

\begin{document}
\baselineskip 18pt%
\begin{titlepage}
\vspace*{1mm}%
\hfill
\vbox{

    \halign{#\hfil         \cr
           } 
      }  
\vspace*{10mm}
\vspace*{12mm}%

\center{ {\bf \Large  Nonlinear Statistical Spline  Smoothers for Critical Spherical Black Hole Solutions in 4-dimension

}}\vspace*{3mm} \centerline{{\Large {\bf  }}}
\vspace*{5mm}
\begin{center}
{ Ehsan Hatefi$^{\dagger,}$$^{\star}$\footnote{E-mails: ehsan.hatefi@uah.es, ehsan.hatefi@sns.it, ahatefi@mun.ca} and  Armin Hatefi$^{\ddagger}$}

\vspace*{0.8cm}{ { \footnotesize
$^{\dagger}$ GRAM Research Group, Department of Signal Theory and Communications, \\
University of Alcala, Alcala de Henares, 28805, Spain.\\
$^{\star}$Scuola Normale Superiore and I.N.F.N,
Piazza dei Cavalieri 7, 56126, Pisa, Italy.\\
$^{\ddagger}$ Department of Mathematics and Statistics, Memorial University of Newfoundland, St John’s, NL, Canada. }}
\vspace*{1.3cm}
\end{center}
\begin{center}{\bf Abstract}\end{center}
\begin{quote}

This paper focuses on self-similar gravitational collapse solutions of the Einstein-axion-dilaton configuration for 
two conjugacy classes of SL(2, R) transformations. These solutions are invariant under spacetime dilation, combined 
with internal transformations. For the first time in Einstein-axion-dilaton literature, we apply the nonlinear statistical 
spline regression methods to estimate the critical spherical black hole solutions in four dimensions. These spline 
methods include truncated power basis, natural cubic spline and penalized B-spline. 
The prediction errors of the statistical models, on average, are almost less
 than $10^{-2}$, so all the developed models can be considered unbiased estimators for the critical collapse functions over their entire domains.
 In addition to this excellence, we derived closed forms and continuously 
differentiable estimators for all the critical collapse functions.

\end{quote}
\end{titlepage}%


\section{Introduction}

 Black holes  are expressed by their mass, angular momentum and their charge. Based on the work of  Choptuik \cite{Chop}, there might be another property that explains the collapse solutions. 
Christodolou \cite{Christodolou} studied in details the spherically symmetric collapse of the scalar field. Choptuik \cite{Chop} showed a critical behaviour of the solutions that demonstrates the discrete self-similarity related to the scalar field gravitational collapse. The solution of gravitational collapse demonstrates spacetime self-similarity in which dilations occur. The critical solution confirms the scaling law. One can express the initial condition of the scalar field by the parameter $p$ corresponding to the field amplitude. Let $p=p_\text{crit}$ define the critical solution. The black hole then forms, when $p$ gets larger than $p_\text{crit}$. When $p$ goes above the critical value, one can determine the mass of black hole or the Schwarzschild radius by the following scaling law
\begin{equation}
r_S(p) \propto M_\text{bh}(p) \propto (p-p_\text{crit})^\gamma\,,
\end{equation}
where  the Choptuik exponent for four dimension and for a real scalar field  is given by $\gamma\simeq 0.37$~\cite{Chop,Hamade:1995ce,Gundlach:2002sx}. However for the other dimensions ($d \geq 4$), all other scalings are given by  \cite{KHA,AlvarezGaume:2006dw}
\begin{equation}
 r_S(p) \propto (p-p_\text{crit})^\gamma \,, \quad M_\text{bh}(p) \sim (p-p_\text{crit})^{(D-3)\gamma}  \,.
\end{equation}
The numerical simulations for other matter content can be found in \cite{Birukou:2002kk,Husain:2002nk,Sorkin:2005vz,Bland:2005vu,HirschmannEardley,Rocha:2018lmv}. Also,  the collapse solutions of the perfect fluid were studied by \cite{AlvarezGaume:2008qs,evanscoleman,KHA,MA}.

 The critical exponent $\gamma \simeq 0.36$ was found 
in \cite{evanscoleman}.  In \cite{Strominger:1993tt}  it is argued that $\gamma$  might have a universal value for all fields that can be coupled to gravity in four dimension.  A nice method to derive critical exponent based on various perturbations of self similar solutions has been introduced in ~\cite{KHA,MA,Hirschmann:1995qx} where one perturbs any field $h$  by the following
\begin{equation}
    h = h_0 + \varepsilon \, h_{-\kappa},
\end{equation}
where $h_{-\kappa}$  has the scaling  $-\kappa \in \mathbb{C}$ related to different modes. The most relevant mode $\kappa^*$ related to the highest value of $\Re(\kappa)$ where the minus sign implied a growing mode near the black-hole formation time $t \rightarrow 0$. It was established in \cite{KHA,MA,Hirschmann:1995qx} that $\kappa^*$ has the following relation with Choptuik exponent  
\begin{equation}
    \gamma = \frac{1}{\Re \kappa^*}\,.
\end{equation}
In \cite{AE} the axial symmetry studied while \cite{AlvarezGaume:2008fx} found  the other critical collapse as well as shock wave. The axion-dilaton in four dimension was first considered in \cite{Hirschmann_1997} and showed that  $\gamma \simeq 0.2641$.

The first motivation of the critical collapse of the axion-dilaton system is the gauge/gravity correspondence \cite{Maldacena:1997re} pointing out the Choptuik exponent, the imaginary part of quasinormal modes and the dual conformal field theory \cite{Birmingham:2001hc}. The second motivation of the axion-dilaton system
 relies on the relationship between the system and formation of black holes, particularly their holographic descriptions in various dimensions \cite{AlvarezGaume:2006dw}. The next motivations  can be the application of the system  to black hole physics \cite{Hatefi:2012bp,Ghodsi_2010}, and the role of S-duality  in this context \cite{Hamade:1995jx}.

 \cite{ours} investigated the whole families of different continuous self similar solutions for the axion-dilaton system in various dimensions
for all conjugacy classes of  SL(2,R). They also extended the results of \cite{AlvarezGaume:2011rk, hatefialvarez1307}.  More importantly,  \cite{Antonelli:2019dqv} carried out the perturbations and regenerated the known value~ \cite{Hirschmann_1997} of $\gamma \sim 0.2641$ in four dimension. The other exponents were derived in \cite{Hatefi:2020gis}. 

In \cite{Hatefi:2021xwh}, we recently explored the idea of polynomial regression and polynomial 
local regression models 
to represent the nonlinear critical collapse functions. 
The polynomial and local regression models can be viewed as polynomial and Fourier series-based 
estimation methods.  Although the polynomial and Fourier basis provide a suitable global estimate 
of the nonlinear function, these methods may result in poor local properties, particularly in 
high noise areas or areas with few observations. In other words, when one uses a polynomial or 
Fourier basis to estimate highly nonlinear critical collapse functions, small changes in a 
parameter of the statistical model will lead to changes in predicted values at every space-time point 
in the domain of the critical collapse functions.  Another challenge associated with local regression methods
is that we have to fit an entire regression model at every point where we want to predict the value 
of the critical collapse function.
All the training data must thus be accessible at every iteration to run the local regression. This 
 challenge may limit the flexibility of the local regression proposals in the cases of large data sets. 

To deal with the challenges of \cite{Hatefi:2021xwh}, we use the property of spline regression
 models to efficiently estimate the non-linear critical collapse functions in elliptic and hyperbolic 
 spaces. As a key set of basis smoothers, 
the spline regression model plays a vital role in estimating the nonlinear function. 
The spline regression methods include truncated power basis regression, natural spline 
 regression and penalized B-spline regression models.

This paper is organized as follows. Section \ref{sec:unperturbed}  describes the  axion-dilaton system in $d$ dimensions
 and three different conjugacy classes of SL(2,R). We present the equations of motions corresponding to the conjugacy 
 classes and solutions in Section \ref{sec:e.m.}. 
The statistical spline  regression methods are explained in Section \ref{sec:stat}. Through numerical studies, we 
evaluate the performance of the proposed spline models in estimating the critical collapse functions in Section \ref{sec:num}. 
Finally, a summary and concluding remarks are presented in Section \ref{sec:con}.

\section{Einstein-axion-dilaton system}\label{sec:unperturbed}

The effective action that describes Einstein-axion-dilaton configuration coupled to  $d$  dimensional gravity is  given by
\begin{equation}
S = \int d^d x \sqrt{-g} \left( R - \frac{1}{2} \frac{ \partial_a \tau
\partial^a \bar{\tau}}{(\Im\tau)^2} \right) \; .
\label{eaction}
\end{equation}
This effective action is described in \cite{Sen:1994fa,Schwarz:1994xn} and the axion-dilaton is defined by $\tau \equiv a + i e^{-\phi}$. 
 The  SL(2,R) symmetry of the effective action is given by
 \begin{equation} \label{eq:sltr}
     \tau \rightarrow M \tau \equiv \frac{a \tau +b}{c \tau + d}, \quad\quad ad-bc=1 
 \end{equation}
 where $a,b,c,d$ are real parameters.
 Based on  the quantum effects, the $SL(2,R)$ symmetry relates to $\mathrm{SL}(2,\mathbb{Z})$. This duality is held even for
the non-perturbative symmetry \cite{gsw,JOE,Font:1990gx}. If we take the variations from the metric and the $\tau$ field, we then derive the equations as \begin{equation}
\label{eq:efes}
R_{ab} = \frac{1}{4 (\Im\tau)^2} ( \partial_a \tau \partial_b
\bar{\tau} + \partial_a \bar{\tau} \partial_b \tau)\,,
\end{equation}
\begin{equation}\label{eq:taueom}
\nabla^a \nabla_a \tau +\frac{ i \nabla^a \tau \nabla_a \tau }{
\Im\tau} = 0 \,.
\end{equation}
From the spherical symmetry, the closed form of the metric $d$ dimensions is
\begin{equation}
    ds^2 = (1+u(t,r)) (-b(t,r)^2 dt^2 + dr^2) + r^2 d\Omega_q^2 \,,
\end{equation}
\begin{equation}
    \tau = \tau(t,r)\, , \quad q \equiv d-2 \; ,
\end{equation}
where 
$d\Omega_q^2$ is the angular part of the metric. One can find  the scale invariant solution as $ (t,r)\rightarrow ( \Lambda t,\Lambda r)$ and $ ds^2 \rightarrow \Lambda^2 ds^2$. 
  Hence all the functions appearing in the metric should be also scale invariant. In other words, they are expressed as $u(t,r) = u(z)$, $b(t,r) = b(z)$, $z \equiv -r/t$. 
The above action is invariant under the SL(2,R) transformation \eqref{eq:sltr} therefore 
$\tau$ must also be invariant and up to an SL(2,R) transformation as follows
\begin{equation}\label{eq:tauscaling}
    \tau(\Lambda t, \Lambda r) = M(\Lambda) \tau(t,r)\,.
\end{equation}
If a system of $(g,\tau)$ meets the above requirements, we then define a continuous self-similar (CSS) solution. Note that different cases do relate to various classes of $\eval{\dv{M}{\Lambda}}_{\Lambda=1}$~\cite{ours}, where $\tau(t,r)$  takes the elliptic, hyperbolic and parabolic assumptions. In the following, we describe the elliptic, hyperbolic and parabolic cases, respectively. 

\subsection{Elliptic class}
The general form of the ansatz for  the elliptic class is given by
\begin{equation}
 \tau(t,r)	=  i \frac{ 1 - (-t)^{i \omega} f(z) }{ 1 + (-t)^{i
\omega} f(z)},
\label{tauansatz_elliptic}
\end{equation}
where $f(z)$ is a complex function satisfying $\abs{f(z)}<1$, and $\omega$ is a real constant. The transformation is a rotation so it shows itself as a symmetry of $f(z)$
\begin{equation}
    f(z) \rightarrow e^{i\theta}f(z)\,,
\end{equation}
 
\subsubsection{Hyperbolic class}

The general form of the ansatz for  the hyperbolic class is
\begin{equation}
    \tau(t,r) = \frac{1-(-t)^w f(z)}{1+(-t)^w f(z)},
    \label{tauansatz_hyperbolic}
\end{equation}
 where one can show that the the ansatz 
 \begin{equation}
\tau(t,r) = (-t)^{\omega} f(z) \,,
\label{tauansatz_hyperbolic66}
\end{equation}
also leads to the same equations of motion for hyperbolic case which is related to \eqref{tauansatz_hyperbolic} by an $\sltr$ transformation where
 $f(z)$ is a complex function satisfying $\Im f(z)>0$, and $\omega$ is a real constant. The compensating transformation is a boost  and the symmetry
is
\begin{equation}
    f(z) \rightarrow e^\lambda f(z)\,,\quad \lambda \in \mathbb{R}.
\end{equation}

\subsubsection{Parabolic class}

The general ansatz  in the parabolic class is shown by
 \begin{equation}
\tau(t,r) = f(z)+\omega \log(-t)\,,
\end{equation}
where $\Im f(z)>0$ and $\omega$ is real and can be taken positive. We have the following residual symmetry
\begin{equation}
    f(z) \rightarrow f(z) + a,
\end{equation}
under which the equations of motion will be invariant.
By replacing the CSS ans\"atze into all the equations of motions, we derive a differential system of equations for $u(z)$, $b(z)$, $f(z)$. By taking the spherical symmetry  $u(z)$ can be expressed in terms of $b(z)$ and $f(z)$
\begin{equation}\label{eq:u0explicit}
    u(z) = \frac{z b'(z)}{(q-1) b(z)}\,,
\end{equation}
 and one can show that all the first derivative of $u(z)$ can be eliminated from all equations of motion  in the elliptic, parabolic and hyperbolic cases as
 \begin{equation}\label{eq:u0pexplicit}
    \frac{qu'(z)}{2(1+u(z))} = \begin{dcases}
    \frac{4zf'(z)\bar f'(z)+2i( w \bar f(z)f'(z) - w f(z)\bar f'(z))}{2(f(z)\bar f(z)-1)^2}  & \\
    \quad\quad\\
    \frac{w  (f'(z) + \bar f'(z)) - 2z  \bar f'(z) f'(z)}{(f(z)-\bar f(z))^2} & \\
     \quad\quad\\
    \frac{w \bar f(z) f'(z)+w  f(z) \bar f'(z) - 2z  \bar f'(z) f'(z)}{(f(z)-\bar f(z))^2} & 
    \end{dcases}
\end{equation}
We finally get the ordinary differential equations (ODEs) as follows 
\begin{align}\label{eq:unperturbedbp}
    b'(z) & = B(b(z),f(z),f'(z))\,, \\
    f''(z) & = F(b(z),f(z),f'(z))\,. \label{eq:unperturbedfpp}
\end{align}
The equations of motion for all three cases are obtained in the following section. 
\section{ The Equations of Motions and the Critical Solutions}\label{sec:e.m.}

The following equations of motion are obtained for self-similar solutions for the elliptic case in any dimension $d = q+2$:
\begin{eqnarray}
b' & = & \frac{ - { 2z(b^2 - z^2)} f' \bar{f}' + {
2i \omega (b^2 - z^2) } (f \bar{f}' - \bar{f} f')
+ {2\omega^2 z |f|^2 }}{q b (1-\abs{f}^2)^2} \,, \nonumber\\
\end{eqnarray}
%
%
%
\begin{eqaed}
q z  \left(z^2-b^2\right) (1-\abs{f}^2)^2 f'' = & \,\, b^2 f'  \big( -2 f \left(q z \bar{f}^2 f'-i \omega z \bar{f}'+q^2 \bar{f}\right) \\ 
& \quad -2 z^2 f' \bar{f}'+2 z \bar{f}  (q-i \omega) f'+q^2 \abs{f}^4+q^2\big)\\
& +z  \Big(2 f^2 \left(q (-1-i \omega) z \bar{f}^2 f' +\omega^2 z \bar{f}' -i q \omega \bar{f} \right)\\
&  \quad + f  \left(2 i \omega z^2 f'  \bar{f}' +2 q z^2 \bar{f}^2 f'^2+4 q z \bar{f}  f' +q \omega (\omega+i)\right)\\
&\quad -2 q z f'  \left(z \bar{f}  f' -i \omega+1\right)-q \omega (\omega-i) \abs{f}^2 f\Big)\\
& +\frac{2 z^3}{b^2} \left(z f' -i \omega f \right)^2 \left(z \bar{f}' +i \omega \bar{f} \right)
\end{eqaed}

The following equations of motion are derived for self-similar solutions in the hyperbolic class, in any dimension $d = q+2$:
%
%
\begin{eqaed}
    b' = -\frac{2 \left(\left(z^2-b^2\right) f' \left(z \bar{f}'-\omega \bar{f}\right)+\omega f \left(\left(b^2-z^2\right) \bar{f}'+\omega z \bar{f}\right)\right)}{q b (f-\bar{f})^2}
\end{eqaed}
\begin{eqaed}
   {q z  \left(z^2-b^2\right) (f-\bar{f})^2} f'' = &\,\, b^2 f' \big(2 z^2 f'\bar{f}'-2 z \omega f \bar{f}'-2\, q\, f (z f'+q \bar{f})\\
   & \quad +2\, q \,z \bar{f} f'+q^2 f^2-2 \omega z \bar{f}{f}'+q^2 \bar{f}^2\big)\\
   +&z \Big(q\omega (1+\omega)f^3-2qz\bar ff'(\bar f-\omega\bar f+zf')\\
   &-2f^2\big(q\omega\bar f+q(1+\omega)zf'-\omega^2z\bar f'\big)\\&
   +f(-q(-1+\omega)\omega\bar f^2+4qz\bar ff'+2z^2f'(qf'-\omega\bar f'))\Big)\\
   &+\frac{2 z^3}{b^2} \left(\omega f-z f'\right)^2(\omega \bar{f}-z \bar{f}')
\end{eqaed}
In all equations, we have five singularities~\cite{AlvarezGaume:2011rk}  at $z = \pm 0$, $z = \infty$ and $z = z_\pm$. The last singularities are expressed by  $b(z_\pm) = \pm z_\pm$ where they are also related to the homothetic horizon where $z=z_+$ is just a coordinate singularity as argued in ~\cite{Hirschmann_1997,AlvarezGaume:2011rk}. Hence by definition $\tau$ must also be regular across it and therefore the $f''(z)$ must be finite as $z\rightarrow z_+$. 
In order to explore more constraints, by vanishing the divergent part of $f''(z)$, we  are able to obtain a complex valued constraint at $z_+$ that can be shown by $G(b(z_+), f(z_+), f'(z_+)) = 0$ and the explicit  forms of $G$  are given by \cite{ours}.

For the elliptic class, $G$ is given by 
\begin{eqaed}\label{eq:Gelliptic}
     G(f(z_+),f'(z_+)) = & \, 2 z \bar{f}(z_+) \left(q^2-2 q-2 \omega^2\right) f'(z_+)\\ & +f(z_+) \bar{f}(z_+) \left(q z_+ \bar{f}(z_+) (-q+2 i \omega+2) f'(z_+)+2 i \omega \left(q+\omega^2\right)\right)\\&-\frac{q z_+ (q+2 i \omega-2) f'(z_+)}{f(z_+)}\\&+q \omega (\omega-i) f(z_+)^2 \bar{f}(z_+)^2-q \omega (\omega+i)\,.
\end{eqaed}
For the hyperbolic case, $G$ is given by
\begin{eqaed}
    G(f(z_+),f'(z_+)) = &\, \bar{f}(z_+) \left(2 z_+ \left(q^2-2 q+2 \omega^2\right) f'(z_+)+q (\omega-1) \omega \bar{f}(z_+)\right)\\& +f(z_+) \left(q z_+ (-q+2 \omega+2) f'(z_+)+2 \omega \bar{f}(z_+) \left(q-\omega^2\right)\right)\\& -\frac{q z_+ \bar{f}(z_+)^2 (q+2 \omega-2) f'(z_+)}{f(z_+)}-q\, \omega (\omega+1) f(z_+)^2\,.
\end{eqaed}
Taking the regularity condition at $z=0$ and considering some residual symmetries, one gets the initial conditions as $b(0) = 1, f'(0) =0$ as well as
\begin{equation}
        f(0) = \left\{\begin{array}{l l l}
        x_0 & \text{elliptic}       & (0<x_0<1) \\
        i x_0 & \text{parabolic} & (0<x_0)\\
        1+i x_0 & \text{hyperbolic} & (0<x_0)
    \end{array}\right.
\end{equation}

where $x_0$ is a real parameter. Therefore we have two distinguished constraints  such as the vanishing of the real and imaginary parts of $G$ and two parameters  $(\omega,x_0)$ to be determined. The solutions were constructed  by numerically integrating in four and five dimensions  and were explored in \cite{Antonelli:2019dqv}. In four dimension and for elliptic case 
the solution was derived in \cite{Eardley:1995ns,AlvarezGaume:2011rk} as
\begin{equation}
    \omega=1.176,\quad \abs{f(0)}=0.892,\quad z_+=2.605 \label{esi}
\end{equation}

 The solutions for four dimensional hyperbolic case were also obtained in \cite{Antonelli:2019dqv}
 and they are called $\alpha$, $\beta$ and $\gamma$ that can be summarized as follows. The $\alpha$-solution is found to be
\begin{equation}
    \omega=1.362,\quad \Im f(0)=0.708,\quad z_+=1.440. \label{esi}
\end{equation}
The $\beta$-solution is given by
\begin{equation}
    \omega=1.003,\quad \Im f(0)=0.0822,\quad z_+=3.29, \label{esi}
\end{equation}
and $\gamma$-solution is is given by
\begin{equation}
    \omega=0.541,\quad \Im f(0)=0.0059,\quad z_+=8.44. \label{esi}
\end{equation}
Note that  we have an extra symmetry  for the parabolic case that is given by
\begin{equation}\label{parabolic_rescaling}
    \omega \rightarrow K \omega\,,\quad f(z) \rightarrow K f(z)\,,\quad K \in \mathbb{R}_+
\end{equation}
and $\tau\rightarrow K \tau$, then $(\omega,\Im f(0))$ as well as $(K \omega,K\Im f(0))$ lead to the same solution. In this case, all equations of motion 
as well as the constraint $G(\omega,\Im f(0))$ are invariant under this scaling.  Further remarks about the parabolic cases can be found in \cite{Hatefi:2020jdr}.

\section{Regression Spline Methods}\label{sec:stat}
The following notations are used to demonstrate spline statistical estimation methods
 in this manuscript.  Let $y$ denote the response variable and ${\bf y}=(y_1,\ldots,y_n)$ 
 represent the vector of responses collected from a random sample of size $n$. Assume we
  have $p$ explanatory variables from the statistical population of interest. 
  Suppose ${\bf X}=({\bf x}_1,\ldots,{\bf x}_p)^\top$ denotes a design matrix of 
  dimensions $ (n \times p)$ representing the realization of $p$ explanatory 
  variables from the random sample of size $n$. Let $\text{rank}({\bf X})=p < n$.

\subsection{Truncated Power Basis Regression}\label{sub:power}
The truncated power basis regression model is the first statistical method that we shall study, in this manuscript, to estimate the 
nonlinear critical collapse functions. The truncated power basis \cite{DeBoor} bridges the gap between the polynomial 
regression model and basis functions.  
Using the higher orders, the truncated power basis combines the advantages of the polynomial
 regression model with the local property of basis functions in the estimation procedure.  

We now describe how the truncated power basis addresses the curvature of a function at ordered points 
$\{l_1 \le l_2 \le \ldots \le l_K\}$ 
(henceforth called knots) in the domain of the functions. Suppose we shall fit two linear polynomial models in a 
neighbourhood of each knot. Accordingly, we fit one polynomial regression into the observations less than knots 
and another for larger observations. Despite this flexibility, a shortcoming of the proposed basis function is 
that the model will not be continuous at each knot $l_k, k=1,\ldots,K$.

The truncated power basis employs a series of constraints and truncated basis functions to guarantee the 
continuity of the developed estimator and the continuous derivatives everywhere in the function's domain, 
including at the knots.  Note that truncated power basis imposes constraints in the estimation procedure.
 These constraints accommodate  
the properties of continuity (of the estimator), the continuity of the first-order derivative, the 
continuity of the second-order derivative and higher orders.  Whenever a constraint is incorporated, 
the degrees of freedom of the developed truncated power basis is reduced by one unit. 

To develop the truncated power basis, we first require to introduce the positive part function as follows
\begin{align}\label{positive}
(x)_+ = \left\{
\begin{array}{lc}
x & x \ge 0\\
0 & x<0.
\end{array}
\right.
\end{align}
Based on the explanatory variable $x$, the truncated power basis \cite{DeBoor} of order $M$ with a set of $K$ knots 
$\{l_1,\ldots,l_K\}$ is given by 
\begin{align}\label{trunc}\nonumber
B_0(x)&=1,\\\nonumber
B_j(x)&=x^j, ~~~ j=1,\ldots,M,\\
B_{M+K}(x)&= (x-l_k)^M_+, ~~~ k=1,\ldots,K.
\end{align}
The truncated power basis \eqref{trunc} entails a polynomial of order $M$ with continuous derivatives of orders  $0$ to $M-1$.
Due to the constraints associated with the assumption of the continuous derivatives, the truncated power basis 
 \eqref{trunc} has $1+M+K$ free parameters to be estimated. 

Given a training data of size $n$ with explanatory variable $x$ and response variable $y$, one can consider $1+M+K$ 
basis functions of \eqref{trunc} as $1+M+K$ explanatory variables in the population.  Hence, the truncated power 
basis \eqref{trunc} now can be viewed as a multiple linear regression model \cite{Harrell,Hatefi:2021xwh} with design matrix ${\bf G}$ of 
dimensions $(n\times (1+M+K))$ whose $(i,j)$-th entry is obtained by
\[
{\bf G}_{i,j} = B_{j-1}(x_i), ~~~ i=1,\ldots,n;~j=1,\ldots,1+M+K. 
\]
Once formulated the method into multiple linear regression and given the design matrix ${\bf G}$, the coefficients 
of the truncated power basis regression model can be estimated by least square method \cite{Harrell,Hatefi:2021xwh} as follows
   \begin{align} \label{ls_trunc}
   {\widehat{\bf\beta}} = \underset{{\bf\beta}}{\arg \min} ~ ||{\bf y} - {\bf G}{\bf\beta}||_2^2,
   \end{align}
  where $||\cdot||_2^2$ denotes the $l_2$ norm. 
  One can easily show the solution to  \eqref{ls_trunc} based on truncated power basis is given by 
  \begin{align}\label{beta_pb}
  {\widehat{\bf\beta}} = ({\bf G}^\top {\bf G})^{-1} {\bf G}^\top {\bf y}.
  \end{align}
  Once the truncated power basis regression model is trained, the response function can be predicted at a new 
  value ${x}_{\text{new}}$ by
  \begin{align} \label{pred_trunc}
  {\widehat y}_{\text{new}} = {\widehat g}({x}_{\text{new}}) = \sum_{j=1}^{1+M+K} {\widehat{\bf\beta}}_j
  B_{j-1}({x}_{\text{new}}),
  \end{align} 
where ${\widehat{\bf\beta}}_j$ is the $j$-th entry of the vector of coefficients 
${\widehat{\bf\beta}}$ from \eqref{beta_pb} and 
$B_{j-1}(\cdot), j=1,\ldots,1+M+K$ are obtained from  \eqref{trunc}. 
From estimation method \eqref{pred_trunc}, one can predict  the nonlinear critical collapse functions at any point 
in their domains. Therefore, the functional forms of the critical collapse solutions can be predicted on the entire 
domain of the  functions in elliptic and hyperbolic spaces. 

Throughout this paper, we focus on the truncated power basis of order $M=3$; henceforth is called a cubic spline.
 Cubic splines are one of the most common spline regression models. This excellence arises from two reasons. 
 First, the human eye can not detect the changes in the third derivative of the underlying function.  
  In addition, the performance merits of the splines rarely increase by the use of higher orders (than $M=3$) 
in the truncated power basis regression models. 

\subsection{Natural Cubic Spline Regression} \label{sub:natural}
The truncated power basis method uses the powers of the polynomial regression to form the basis 
expansions in the spline regression. Accordingly, the truncated power basis performance changes 
erratically in estimating the nonlinear critical collapse functions over the extreme regions. 
The extreme regions include the space-time points, which are less than the lowest knot $l_1$ 
and greater than the largest knot $l_K$ in the domain of the functions. Natural spline regression 
is a solution to this estimation challenge. 

On top of the constraints required for the truncated power basis method, the natural spline method 
imposes one more constraint to control the erratic changes of the truncated power basis. The natural spline 
 model of order $M$ shrinks the power basis function to an order of $(M-1)/2$ in the extreme regions. 
  Due to the advantages of the cubic splines (i.e. $M=3$), we only focus on the natural splines of 
 order M=3, henceforth called natural cubic splines. Thus, the natural cubic splines propose a linear 
 estimator in the extreme regions beyond the extremal knots to control the erratic changes in 
 estimating nonlinear critical collapse functions.  
 
 Given a training sample of size $n$ with explanatory variable $x$ and response variable $y$, 
 the natural cubic splines \cite{DeBoor} based the knots $\{l_1,\ldots,l_K\}$ first constructs the basis functions as follows
\begin{align}\label{natural}\nonumber
B_0(x)&=1,\\\nonumber
B_j(x)&=x,\\
B_{1+j}(x)&= d_j(x)-d_{K-1}(x), ~~~ j=1,\ldots,l_{K-2},
\end{align}
where
\[
d_j(x) = (x-l_j)^3_+ - (x-l_K)^3_+ / (l_K-l_j),
\]
with $(\cdot)_+$ is given by \eqref{positive}.

 Once the basis functions were formed, 
the natural cubic spline regression can be treated as a multiple linear regression model with design matrix ${\bf G}$  
of dimensions $(n\times K)$ whose $(i,j)$-th entry is calculated by
\begin{align}\label{G_nat}
{\bf G}_{i,j} = B_{j-1}(x_i), ~~~ i=1,\ldots,n;~j=1,\ldots,K. 
\end{align}
where $B_{j(\cdot)}$ is obtained from \eqref{natural}.  
Now similar to the truncated power basis spline, the cubic natural spline estimate of the critical collapse function $g(x)$
at a new value ${x}_{\text{new}}$ is given by
  \begin{align} \label{pred_natural}
  {\widehat y}_{\text{new}} = {\widehat g}({x}_{\text{new}}) = \sum_{j=1}^{K} {\widehat{\bf\beta}}_j
  B_{j-1}({x}_{\text{new}}).
  \end{align} 
The vector of coefficients estimate 
${\widehat{\bf\beta}}$ is computed by 
  ${\widehat{\bf\beta}} = ({\bf G}^\top {\bf G})^{-1} {\bf G}^\top {\bf y}$
where design matrix ${\bf G}$ is obtained from the natural cubic basis from \eqref{G_nat}.

\subsection{Penalized B-Spline Regression} \label{sub:pen}
The polynomial and local regression models \cite{Hatefi:2021xwh} estimate the global pattern of 
the critical collapse functions. 
Small local changes impact the global behaviour of the proposed estimates. The truncated power basis 
spline and natural spline models accommodate the local changes in the prediction using the basis
 functions. At the same time, we do not require to adjust the global pattern of the estimated 
 critical collapse functions. The truncated power basis and natural splines exploit knots over 
 the domain of the function to handle the local behaviour of the estimates.   

 Despite the knots in the truncated power basis and natural splines, we observe that the basis functions
  are nonzero at the $k$-th knot for all data points greater than the knot. Also, the columns of the 
  design matrix corresponding to the higher-order terms are almost nonzero for all data points. The 
  local property is allowed in the truncated power basis and natural splines at the price of a dense 
  design matrix and collinearity challenge. In other words, the interaction between knot-based functions
   in truncated power basis and natural splines lead to a dense design matrix whose elements are almost 
   nonzero everywhere, even in the case of a small data set. In addition, the columns of the design matrix 
   depend linearly on each other. These challenges make the design matrix ill-conditioned in the cases of 
   large data sets and a large set of knots 
 in the estimation of critical collapse functions. In this subsection, we study the idea of B-spline regression
  models to handle the above challenges.  
 
The B-spline basis functions exploit the idea of the truncated power basis to allow the local property to develop 
diagonal bands on the designs matrix to avoid the above challenges. Given a training data of size $n$ from explanatory 
variable $x$ and response variable $y$, 
 the B-spline regression model \cite{DeBoor,Eilers} of order $M$ with knots $\{l_1,\ldots,l_K\}$ introduces the basis expansions as follows:
 \begin{align*}
B_{j,0}(x)&=
\left\{
\begin{array}{lc}
1 &  l_j \le x \le l_{j+1},\\
0 & \text{otherwise,}
\end{array}
\right.
\end{align*}
 and then we compute $B_{j,m}(x), j=1,\ldots,K$ and $m=1,\ldots,M$ recursively as
\begin{align}\label{bspline}
B_{j,m}(x)=& \frac{x - l_j}{l_{j+m} - l_j} B_{j,m-1}(x) 
          + \frac{l_{j+m+1} - x}{l_{j+m+1} - l_{j+1}} B_{j+1,m-1}(x). 
\end{align}
B-splines make use of local support property in the prediction problem. First, the design matrix columns 
are non-zero for only a small subset of points in the regions specified by the knots. Also, the B-spline 
replaces the polynomial basis with step functions (for $M=1$) and symmetric 
 triangular (for $M=2$) to catch the non-linear functions for every region between the knots.  Similar 
 patterns are observed for higher orders of B-splines. The B-spline regression model can be viewed as a
  multiple regression model with design matrix ${\bf G}$ of dimensions $(n\times(K+M))$ whose columns are obtained by 
 \begin{align}\label{G_bspline}
 G_{i,j+m} = B_{j,m}(x_i), ~~ j=1,\ldots,K;m=1,\ldots,M;i=1,\ldots,n,
 \end{align}
 where $B_{j,m}(\cdot)$ comes from \eqref{bspline}. Now the critical collapse function $g(x)$ can be predicted 
 at a new point ${x}_{\text{new}}$ by 
   \begin{align} \label{pred_bspline}
  {\widehat y}_{\text{new}} = {\widehat g}({x}_{\text{new}}) = \sum_{j=1}^{K+M} {\widehat{\bf\beta}}_j
  B_{j}({x}_{\text{new}}),
  \end{align} 
 where vector of coefficients estimate is given by ${\widehat{\bf\beta}} = ({\bf G}^\top {\bf G})^{-1} {\bf G}^\top {\bf y}$
 with $B_{j}(\cdot)$ and design matrix ${\bf G}$ are computed from \eqref{bspline} and \eqref{G_bspline}, respectively.
 
The B-spline proposal requires a set of knots on the domain of the nonlinear critical collapse functions. 
The next question can be the optimal number of knots required in B-spline smoother to reach an optimized 
amount of smoothness in the estimation and avoid overfitting problems. The penalized B-spline regression model 
is a solution to this problem. Penalized B-splines are designed to estimate the regression coefficients 
$\beta_1,\ldots,\beta_K$ in \eqref{pred_bspline} under the constraints to optimize the level of smoothness 
and avoid overfitting. The optimized can be found through some penalty terms. There are various options for 
the penalty term; however, the most common one is the $l_2$ penalty where we choose threshold $t$ such that 
$\sum_{j=1}^{K} \beta_j^2 \le t$. 
 \begin{figure}
\includegraphics[width=1\textwidth,center]{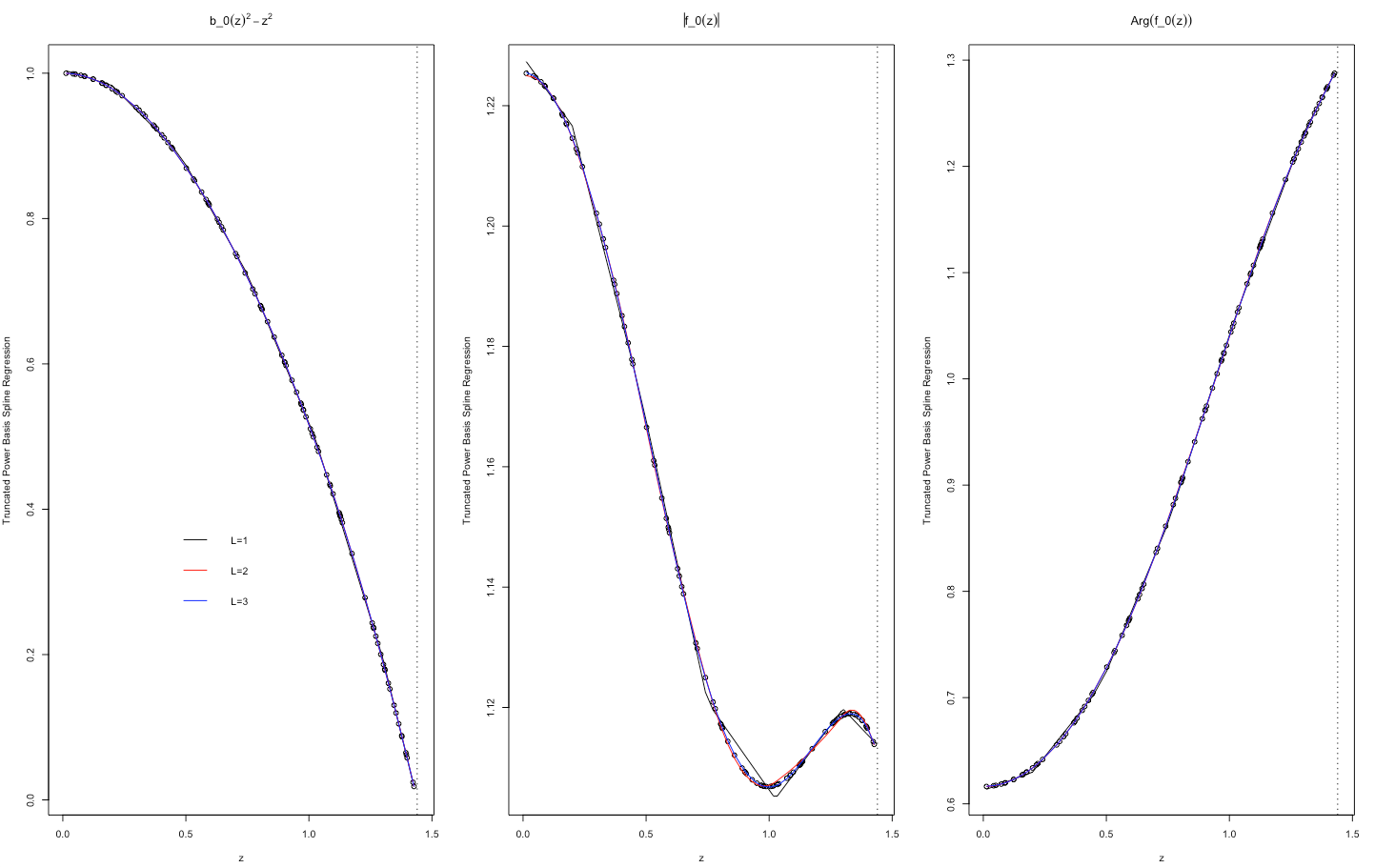}
\caption{\footnotesize{The estimates of critical collapse functions corresponding to $\alpha$-solution of hyperbolic case based on truncated power basis  
regression of orders $l=\{1,2,3\}$ from a training sample of size $n=100$.}}
 \label{hy_al_trunc}
\end{figure}
  
  Given a training data of size $n$ from explanatory variable $x$, response variable $y$ and a large set of knots $\{l_1,\ldots,l_K\}$,
 we first obtain the B-spline basis from \eqref{bspline} and construct the corresponding design matrix ${\bf G}$ from \eqref{G_bspline}. 
 The penalized B-spline estimation can then be re-written as penalized least square method as follows
   \begin{align}\label{pls}
   {\widehat{\bf\beta}} = \underset{{\bf\beta}}{\arg \min} \left\{ ||{\bf y} - {\bf G}{\bf\beta}||_2^2 + 
   \lambda^2 {\bf\beta}^\top {\bf D} {\bf\beta} \right\},
   \end{align}
  where $\lambda$ is the tuning parameter and
  \[
  {\bf D}= 
  \left[
  \begin{array}{cc}
  {\bf 0}_{2\times 2} & {\bf 0}_{2\times K} \\
  {\bf 0}_{K\times 2} & {\bf I}_{K\times K}
  \end{array}
  \right].
  \]
  From optimization \eqref{pls}, it is easy to show that the penalize B-spline estimates the critical 
  collapse function $g(x)$ at new ${x}_{\text{new}}$  is obtained by
   \begin{align} \label{pred_penalized}
  {\widehat y}_{\text{new}} = {\widehat g}({x}_{\text{new}}) = 
  {\bf B}({x}_{\text{new}}) ({\bf G}^\top {\bf G}+ \lambda {\bf D})^{-1} {\bf G}^\top {\bf y}
    \end{align}
  where  ${\bf B}({x}_{\text{new}})=
  \left(
  1, B_1({x}_{\text{new}}),\ldots, B_K({x}_{\text{new}})
  \right)^\top$ and $B_j({x}_{\text{new}}), j=1,\ldots,K$ are calculated from \eqref{bspline}.

\begin{figure}
\includegraphics[width=1\textwidth,center]{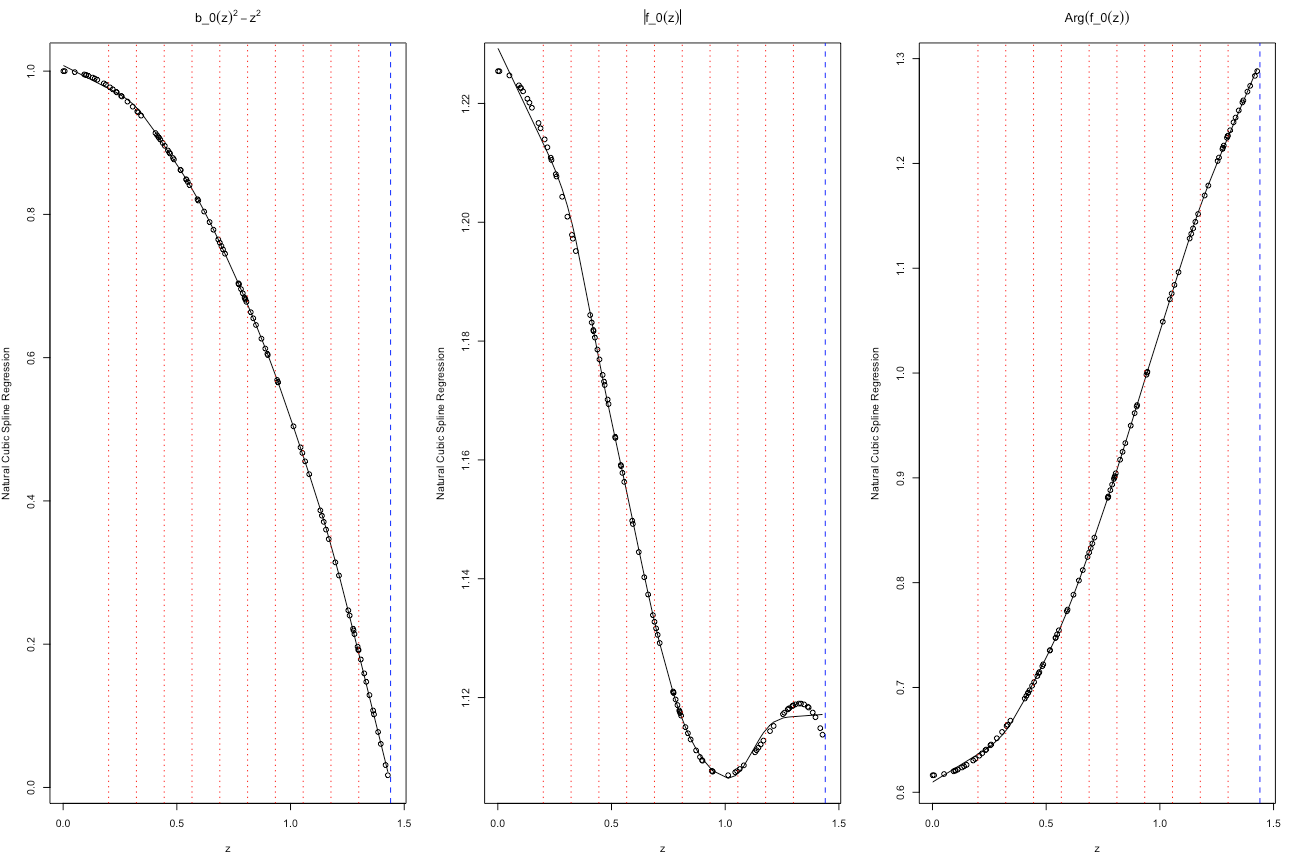}
\caption{\footnotesize{The estimates of critical collapse functions corresponding to $\alpha$-solution of 
hyperbolic case based on the natural cubic spline 
regression from a training sample of size $n=100$. The  red lines show the locations of the knots. }}
 \label{hy_al_nat}
\end{figure}

\begin{table}[ht]
\centering
\footnotesize{\begin{tabular}{lccccc}
  \hline
  $K$   &   Elliptic &            & Hyperbolic &  \\ 
  \cline{3-5} 
        &            & $\alpha$-solution   & $\beta$-solution    & $\gamma$-solution \\ 
\hline

 3 & 0.0028825 & 0.0024568 & 0.0031036 & 0.0000290 \\ 
 5 & 0.0013078 & 0.0000938 & 0.0004437 & 0.0000107 \\ 
 8 & 0.0002640 & 0.0000104 & 0.0000209 & 0.0000110 \\ 
 10 & 0.0000988 & 0.0000056 & 0.0000268 & 0.0000113 \\ 
 15 & 0.0001045 & 0.0000055 & 0.0000149 & 0.0000119 \\ 
 20 & 0.0000552 & 0.0000050 & 0.0000057 & 0.0000129 \\ 
   \hline
\end{tabular}
\caption{\footnotesize{ The $\sqrt{\text{MSE}}$ of truncated power basis method of orders $M=3$ with $K=\{3,5,\ldots,20\}$ knots in estimating 
the critical collapse response function $g(z)=|f_0(z)|$ in elliptic and hyperbolic cases based on a training sample of size $n=100$.}}
\label{absf_truncated}}
\end{table}

As we see from \eqref{pred_penalized}, the penalized B-spline estimator requires us to determine the tuning
 parameter $\lambda$ in estimating the critical collapse function. We apply the idea of cross-validation 
 to find the optimal value of $\lambda$. The idea of cross-validation arises from the fact that the optimal 
 $\lambda$ provides the best fit to the underlying nonlinear function over the entire domain. Let $y_i$ 
 and ${\hat y}_i$ be the observed and predicted responses using $\lambda$ for all training data $i=1,\ldots,n$. 
 To this end, we find the optimal $\lambda$ value, which minimizes the error sum of squares as a measure of
  goodness of fit. The error sum of squares (RSS) based on each $\lambda$ is given by   
   \[
   \text{RSS}(\lambda) = \sum_{i=1}^{n} (y_i - {\hat y}_i)^2 = {\bf y}^\top {\bf y} - 2 {\bf y}^\top 
   {\bf \widehat g}_\lambda + {\bf \widehat g}_\lambda^\top {\bf \widehat g}_\lambda,
   \] 
    where ${\bf y}=(y_1,\ldots,y_n)$ and ${\bf \widehat g}_\lambda = \left({\bf \widehat g}_\lambda(x_1),
    \ldots,{\bf \widehat g}_\lambda(x_n)\right)$ are vectors of the observed and the predicted responses
     based on full chain of training data for each $\lambda$.

  Generalized cross-validation (GCV) \cite{Eilers} is considered as one of the well-established algorithms to find
   the optimal $\lambda$ that dramatically reduces the computational burden of the cross-validation. 
   The GCV makes use of the Demmler--Reinsch Orthogonalization approach \cite{Ruppert} to find sequentially the predicted responses for each $\lambda$ value. Accordingly, the GCV procedure is carried out as follows: First, we predict the value of nonlinear function ${\bf \widehat g}_\lambda$ based on all $x$ values in the training data for each $\lambda$ as
   \[
   {\bf \widehat g}_\lambda = {\bf G} \left( {\bf G}^\top{\bf G} +\lambda {\bf D}\right)^{-1} {\bf G}^\top {\bf y}.
   \]
   
\begin{table}[ht]
\centering
\footnotesize{\begin{tabular}{lccccc}
  \hline
  $K$   &   Elliptic &            & Hyperbolic &  \\ 
  \cline{3-5} 
        &            & $\alpha$-solution   & $\beta$-solution    & $\gamma$-solution \\ 
\hline
 3 & 0.0120394 & 0.0080850 & 0.0073040 & 0.0015161 \\ 
 5 & 0.0047826 & 0.0019024 & 0.0048836 & 0.0005355 \\ 
 8 & 0.0044198 & 0.0011526 & 0.0024039 & 0.0002743 \\ 
 10 & 0.0036682 & 0.0009570 & 0.0016642 & 0.0002136 \\ 
 15 & 0.0023819 & 0.0007245 & 0.0009556 & 0.0001528 \\ 
 20 & 0.0019741 & 0.0005985 & 0.0005809 & 0.0001225 \\ 
   \hline
\end{tabular}
\caption{\footnotesize{ The $\sqrt{\text{MSE}}$ of cubic natural spline method  with $K=\{3,5,\ldots,20\}$ knots in estimating 
the critical collapse response function $g(z)=|f_0(z)|$ in elliptic and hyperbolic cases based on a training sample of size $n=100$.}}
\label{absf_natural}}
\end{table}

    In the second step, we apply Cholesky decomposition for symmetric matrix ${\bf G}^\top{\bf G}$ and obtain the triangular matrix ${\bf R}$ such that 
    ${\bf G}^\top{\bf G} = {\bf R}^\top{\bf R}$. In step three,  we apply singular value decomposition  of  the symmetric matrix ${\bf R}^{-1} {\bf D} {\bf R}$ and find
     ${\bf U}\text{diag}({\bf s}) {\bf U}^\top$. We compute ${\bf A}={\bf G}{\bf R}^{-1}{\bf U}$ and ${\bf b}= {\bf A}^\top{\bf y}$.
     Then, ${\bf \widehat g}_\lambda = \left({\bf \widehat g}_\lambda(x_1),
    \ldots,{\bf \widehat g}_\lambda(x_n)\right)$ is updated by
    \[
    {\bf \widehat g}_\lambda = {\bf A} \left( \frac{{\bf b}}{1+\lambda {\bf s}}\right).
    \]
    We  perform the above procedure sequentially  for all possible $\lambda$ values and obtain the RSS based on all $\lambda$ values; that is 
    RSS($\lambda_1$), RSS($\lambda_2$) and so on. Finally, the optimal value is the $\lambda$ corresponding to the minimum RSS. This optimal value is henceforth called $\lambda_{\text{GCV}}$. From  $\lambda_{\text{GCV}}$ and  \eqref{pred_penalized}, 
    the penalized B-spline estimator for nonlinear function at the new point ${x}_{\text{new}}$ is derived by
      \begin{align} \label{pred_gcv}
  {\widehat y}_{\text{new}} = {\widehat g}({x}_{\text{new}}) = 
  {\bf B}({x}_{\text{new}}) ({\bf G}^\top {\bf G}+ \lambda_{\text{GCV}} {\bf D})^{-1} {\bf G}^\top {\bf y}.
    \end{align}

\begin{figure}
\includegraphics[width=1\textwidth,center]{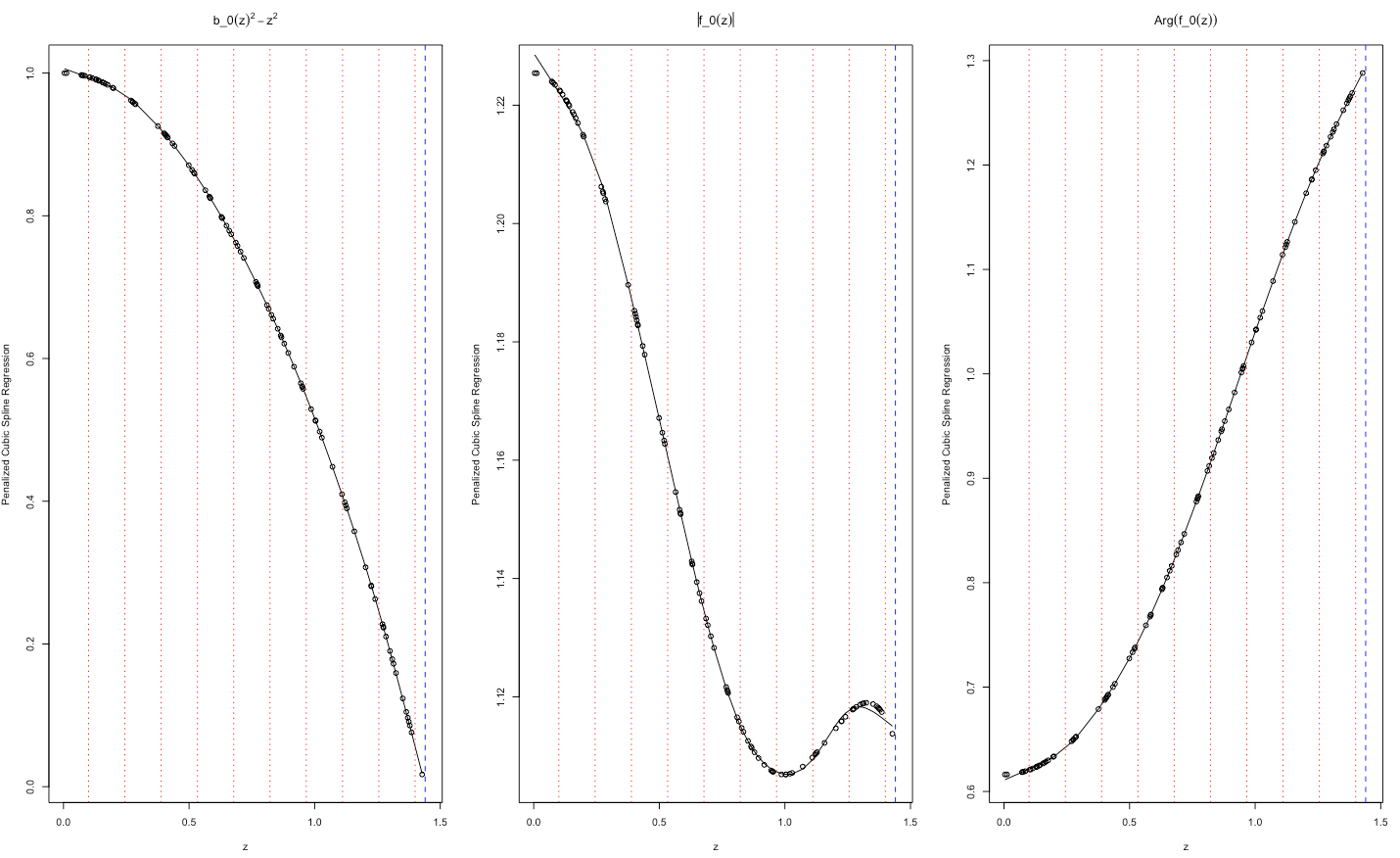}
\caption{\footnotesize{The estimates of critical collapse functions corresponding to $\alpha$-solution of hyperbolic case based on penalized B-spline
regression of order $l=1$ from a training sample of size $n=100$. The  red lines show the locations of the knots.}}
 \label{hy_al_pen}
\end{figure}

\section{Numerical Studies} \label{sec:num}
Solutions to axion-dilaton  in black holes were recently studied in \cite{ours} in both elliptic and hyperbolic spacetime in 4 dimensions. 
  They located only one solution in the elliptic case and three distinctive solutions
   in the hyperbolic case. Due to the importance of the unperturbed critical collapse 
   functions in the location of the solutions, in \cite{Hatefi:2021xwh}  we recently utilized polynomial regression, kernel regression 
 and local regression models to estimate the unperturbed critical function. While these 
 polynomial and Fourier-based methods work well in estimating the global behaviour of 
 the nonlinear critical collapse functions, these estimators do not support local properties.
  This section develops truncated power basis spline, cubic natural spline, and penalized 
  B-spline models in estimating the nonlinear critical collapse functions in elliptic and hyperbolic cases.

In a similar vein to \cite{ours}, we implemented a numerical optimization to find the critical solutions 
to the equations of motions in the domain of forward-singularity ($[0,z_+]$). This resulted in a unique solution 
 in the interval [0, 2.5] corresponding to the domain of the critical collapse functions in the elliptic space.  
 In the case of hyperbolic space, we found three solutions corresponding to $\alpha$-, 
$\beta$- and $\gamma$- solutions to the equations of motions.  
This leads to three domains for the critical collapse functions corresponding to interval [0,1.44] (for $\alpha$-solution), 
interval [0, 3.30] (for $\beta$-solution) and interval [0, 8.45] (for $\gamma$-solution). Similar to \cite{ours,Hatefi:2021xwh}, 
we implemented the optimization search and obtained 2000 observations from each domain of the critical collapse functions 
$b_0^2-z^2,|f_0(z)|$ and $\arg f_0(z)$ in both elliptic ad hyperbolic spaces.  These 2000 observations were
 then treated as our unknown statistical populations that should be recovered by the statistical models for 
 each interval in elliptic and hyperbolic spaces. 
 
From each observation in the population, we treated the spacetime $z$ coordinate and the value of the nonlinear
 critical collapse functions as the explanatory variable and response received from the population. 
We selected a random sample (with replacement) of size $n=100$ from each population of interest for training the 
statistical models. We then use the training data to estimate the parameters of the truncated power basis spline,
 cubic natural spline and penalized B-spline models for estimating the three critical collapse functions. To 
 assess the performance of the proposed spline regression models, independently from the training data, we also 
 generated a validation/test sample of size $n=100$ from each population. Suppose 
 $(x_{i,test},y_{i,test}), i=1\ldots,n$ denotes the test sample of size $n$.  Using the trained truncated 
 power basis, cubic natural spline and penalized B-spline estimators, we predicted the critical collapse 
 functions at every point $x_{i,test}$ (without using the values of $y_{i,test}$ in the estimation procedure)
  $i=1\ldots,n$. The predicted response is called ${\widehat y}_{i,test}, i=1\ldots,n$. We then computed the 
  prediction responses as described above for all critical collapse functions and used the square root of 
  the mean squared error to measure 
 the prediction accuracy of the purposed spline estimation methods. 
 The square root of mean squared error $\sqrt{\text{MSE}}$
is given by 
\[
\sqrt{\text{MSE}} = 
\left(
1/n \sum_{i=1}^{n} 
({\widehat y}_{i,test} - y_{i,test})^2
\right)^{1/2}.
\]
$\sqrt{\text{MSE}}$ decreases when the proposed spline model predicts more accurately the underlying nonlinear critical 
collapse function over the entire domain. In other words, the predicted responses (using the spline model) get 
closer to the responses observed from the nonlinear critical collapse function.

\begin{table}[ht]
\centering
\footnotesize{\begin{tabular}{lccccc}
  \hline
  $K$   &   Elliptic &            & Hyperbolic &  \\ 
  \cline{3-5} 
        &            & $\alpha$-solution   & $\beta$-solution    & $\gamma$-solution \\ 
\hline
 3 & 0.0111553 & 0.0073981 & 0.0071837 & 0.0018790 \\ 
 5 & 0.0028832 & 0.0018389 & 0.0039828 & 0.0005865 \\ 
 8 & 0.0025321 & 0.0009167 & 0.0016816 & 0.0003127 \\ 
 10 & 0.0019106 & 0.0007042 & 0.0011894 & 0.0002366 \\ 
 15 & 0.0015122 & 0.0004346 & 0.0006268 & 0.0001317 \\ 
 20 & 0.0011401 & 0.0003069 & 0.0003900 & 0.0001016 \\ 
   \hline
\end{tabular}
\caption{\footnotesize{ The $\sqrt{\text{MSE}}$ of penalized cubic B-spline method  with $K=\{3,5,\ldots,20\}$ knots in estimating 
the critical collapse response function $g(z)=|f_0(z)|$ in elliptic and hyperbolic cases based on a training sample of size $n=100$.}}
\label{absf_penalized}}
\end{table}

\begin{figure}
\includegraphics[width=1.2\textwidth,center]{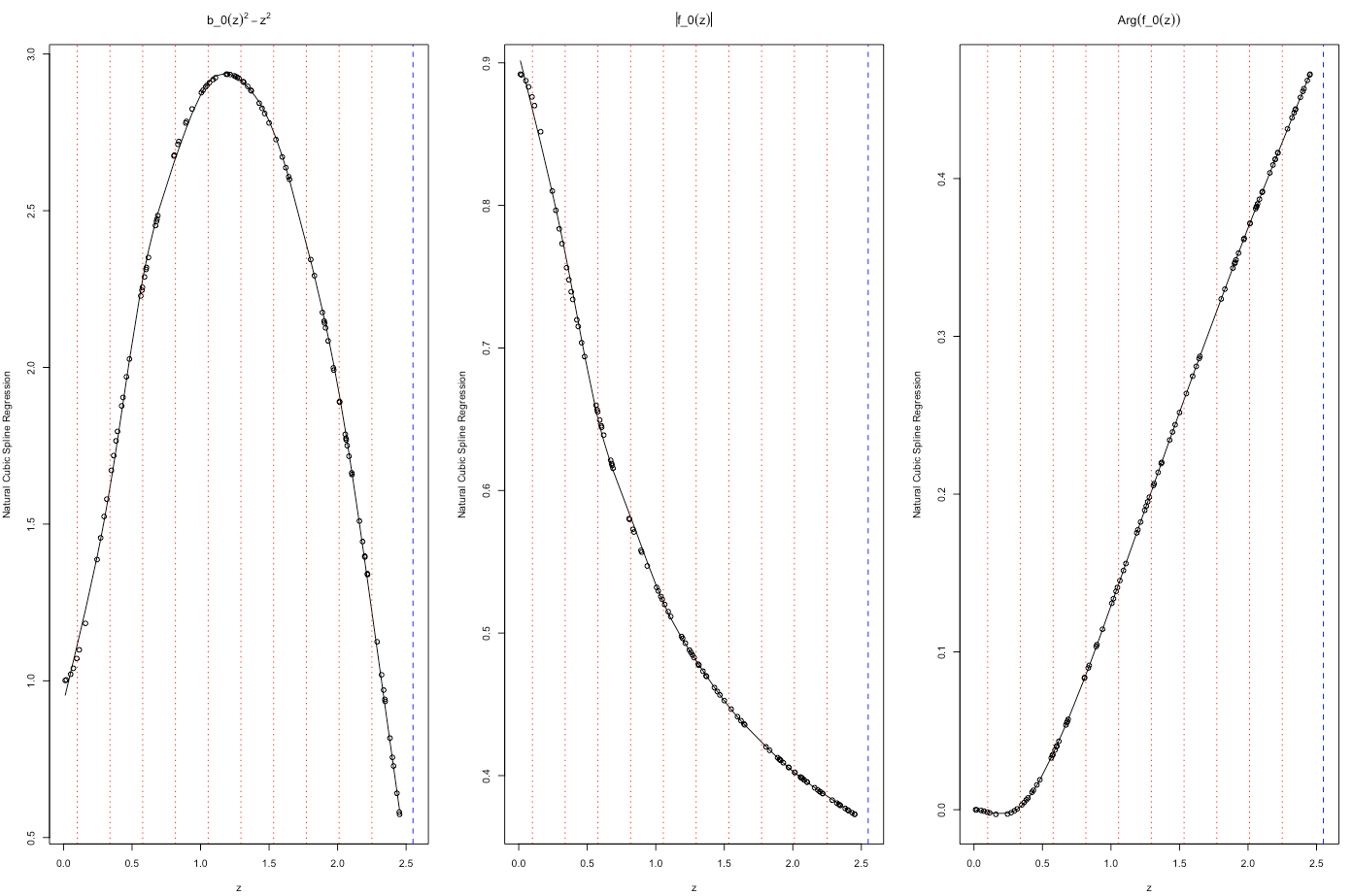}
\caption{\footnotesize{The estimates of critical collapse functions based on the natural cubic spline
regression in elliptic case from a training sample of size $n=100$. The  red lines show the locations of the knots.}}
 \label{el_nat}
\end{figure}

\begin{figure}
\includegraphics[width=1.2\textwidth,center]{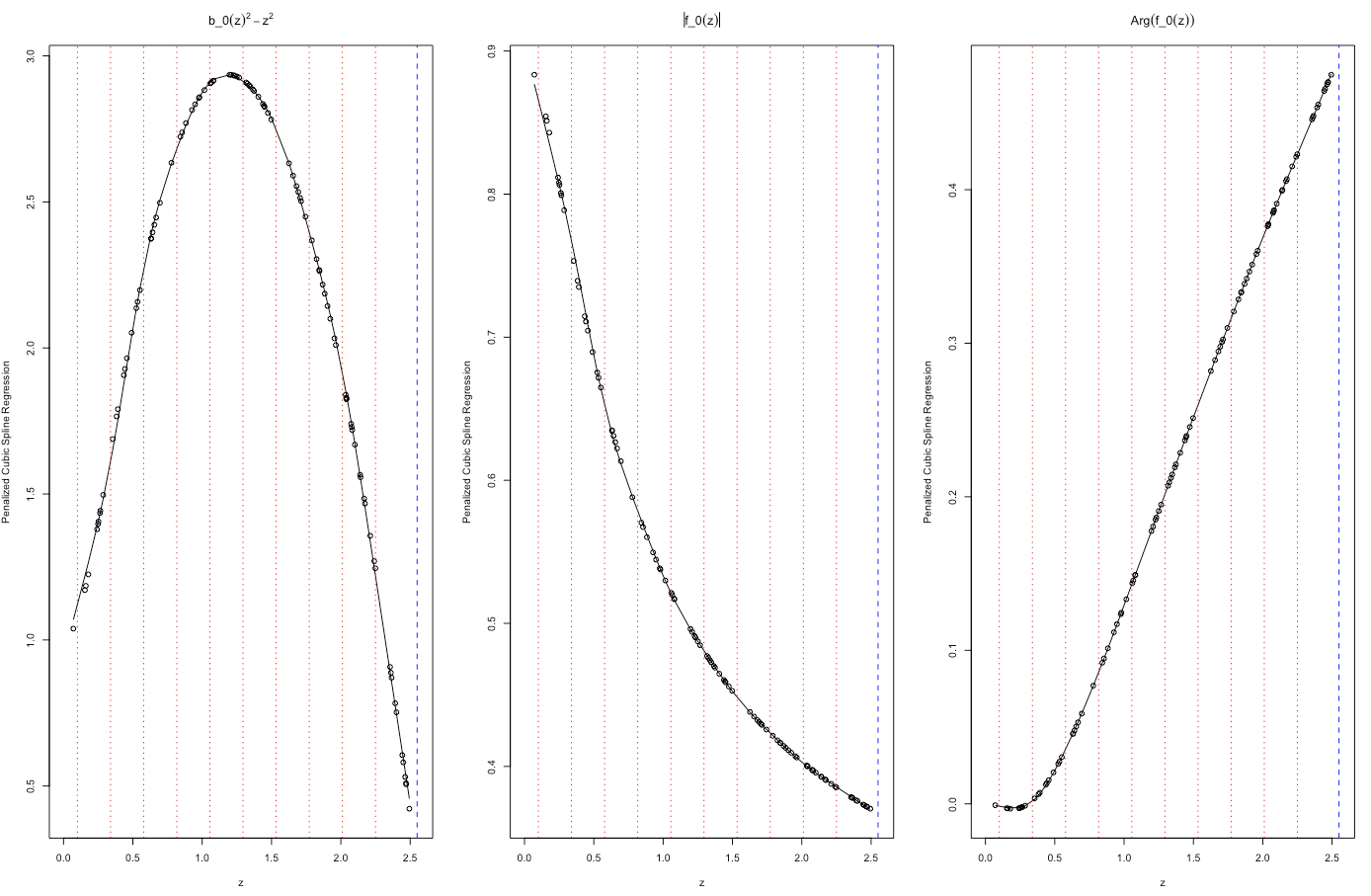}
\caption{\footnotesize{The estimates of critical collapse functions based on penalized B-spline
regression of order $l=1$ in elliptic from a training sample of size $n=100$. The  red lines show the locations of the knots.}}
 \label{el_pen}
\end{figure}

\begin{figure}
\includegraphics[width=1.2\textwidth,center]{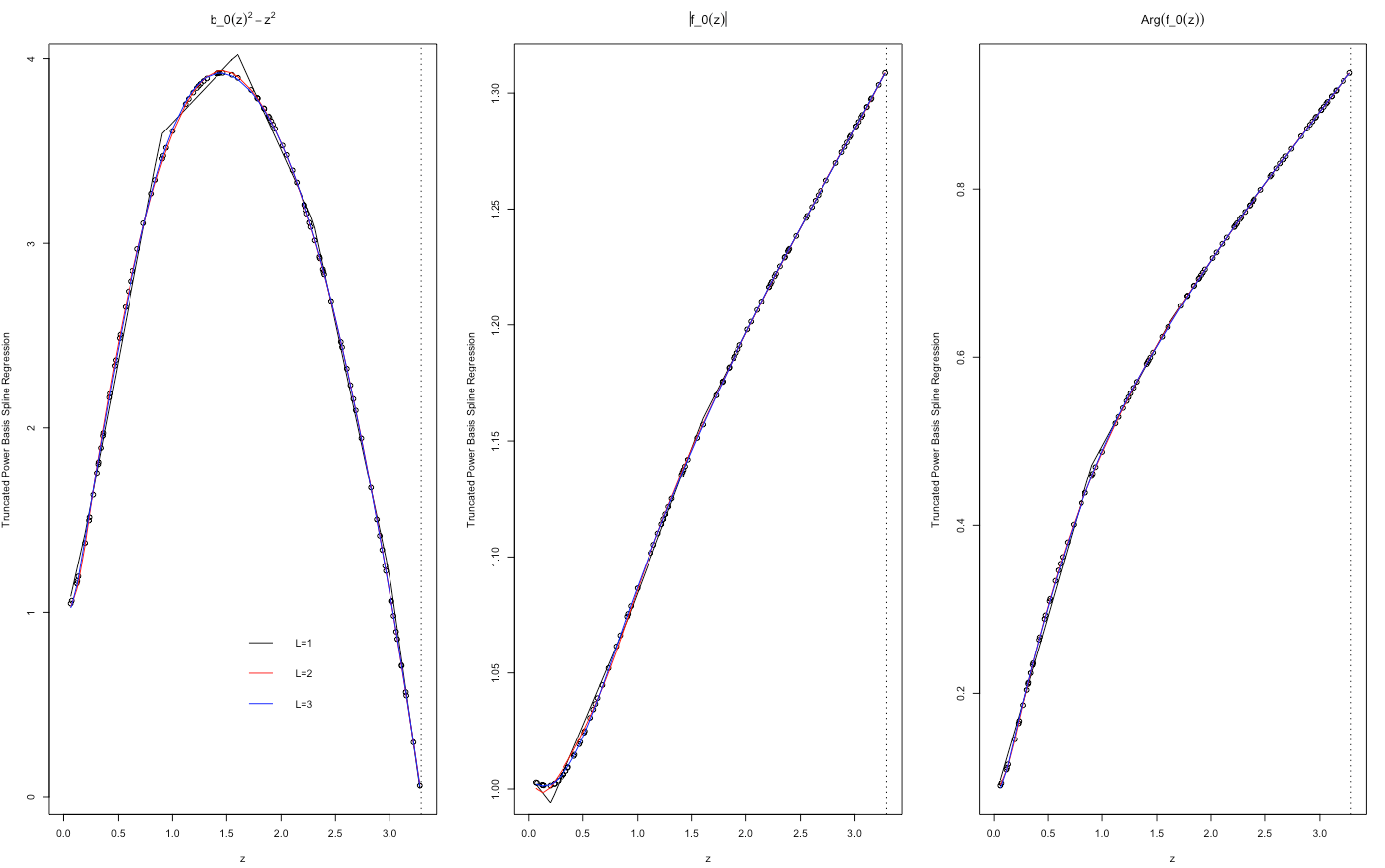}
\caption{\footnotesize{The estimates of critical collapse functions corresponding to $\beta$-solution of hyperbolic case based on truncated power basis  
regression of orders $l=\{1,2,3\}$ from a training sample of size $n=100$.}}
 \label{hy_be_trunc}
\end{figure}

\begin{figure}
\includegraphics[width=1.2\textwidth,center]{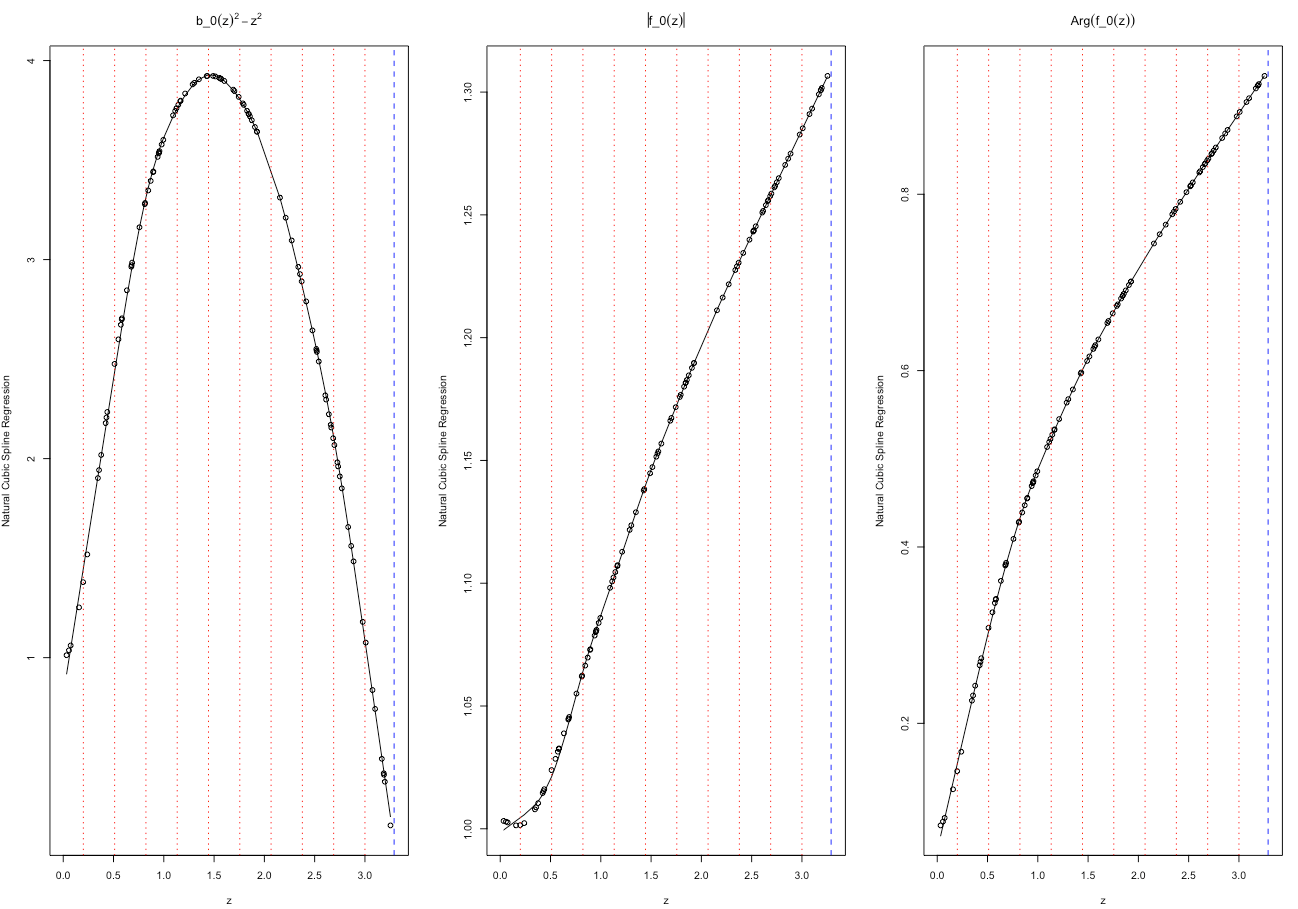}
\caption{\footnotesize{The estimates of critical collapse functions corresponding to $\beta$-solution of hyperbolic case based on natural cubic spline 
regression from a training sample of size $n=100$. The  red lines show the locations of the knots.}}
 \label{hy_be_nat}
\end{figure}

\begin{figure}
\includegraphics[width=1.2\textwidth,center]{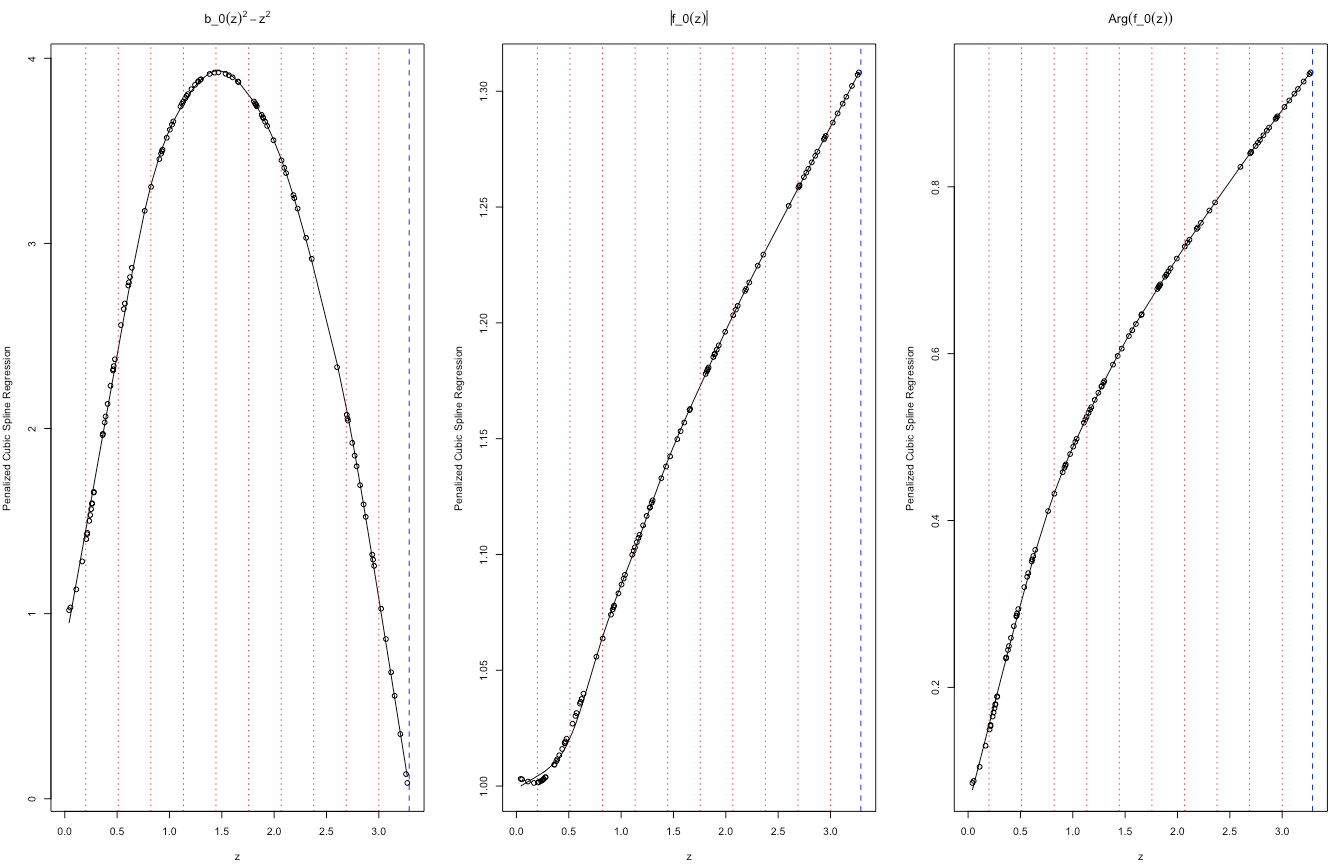}
\caption{\footnotesize{The estimates of critical collapse functions corresponding to $\beta$-solution of hyperbolic case based on penalized B-spline  
regression of order $l=1$ from a training sample of size $n=100$.The  red lines show the locations of the knots.}}
 \label{hy_be_pen}
\end{figure}

\begin{figure}
\includegraphics[width=1.2\textwidth,center]{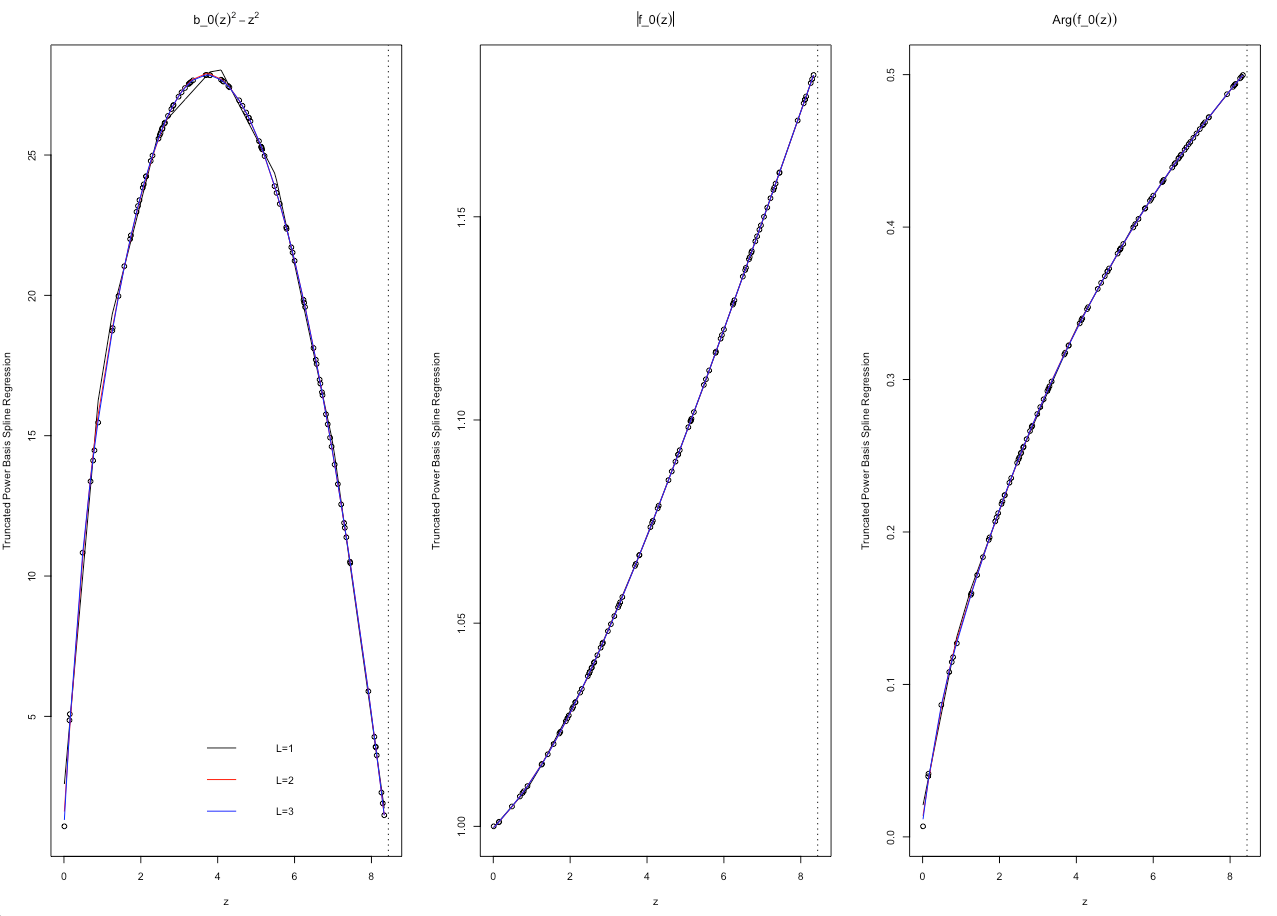}
\caption{\footnotesize{The estimates of critical collapse functions corresponding to $\gamma$-solution of hyperbolic case based on truncated power basis 
regression  of orders $l=\{1,2,3\}$ from a training sample of size $n=100$.}}
 \label{hy_ga_trunc}
\end{figure}

\begin{figure}
\includegraphics[width=1.2\textwidth,center]{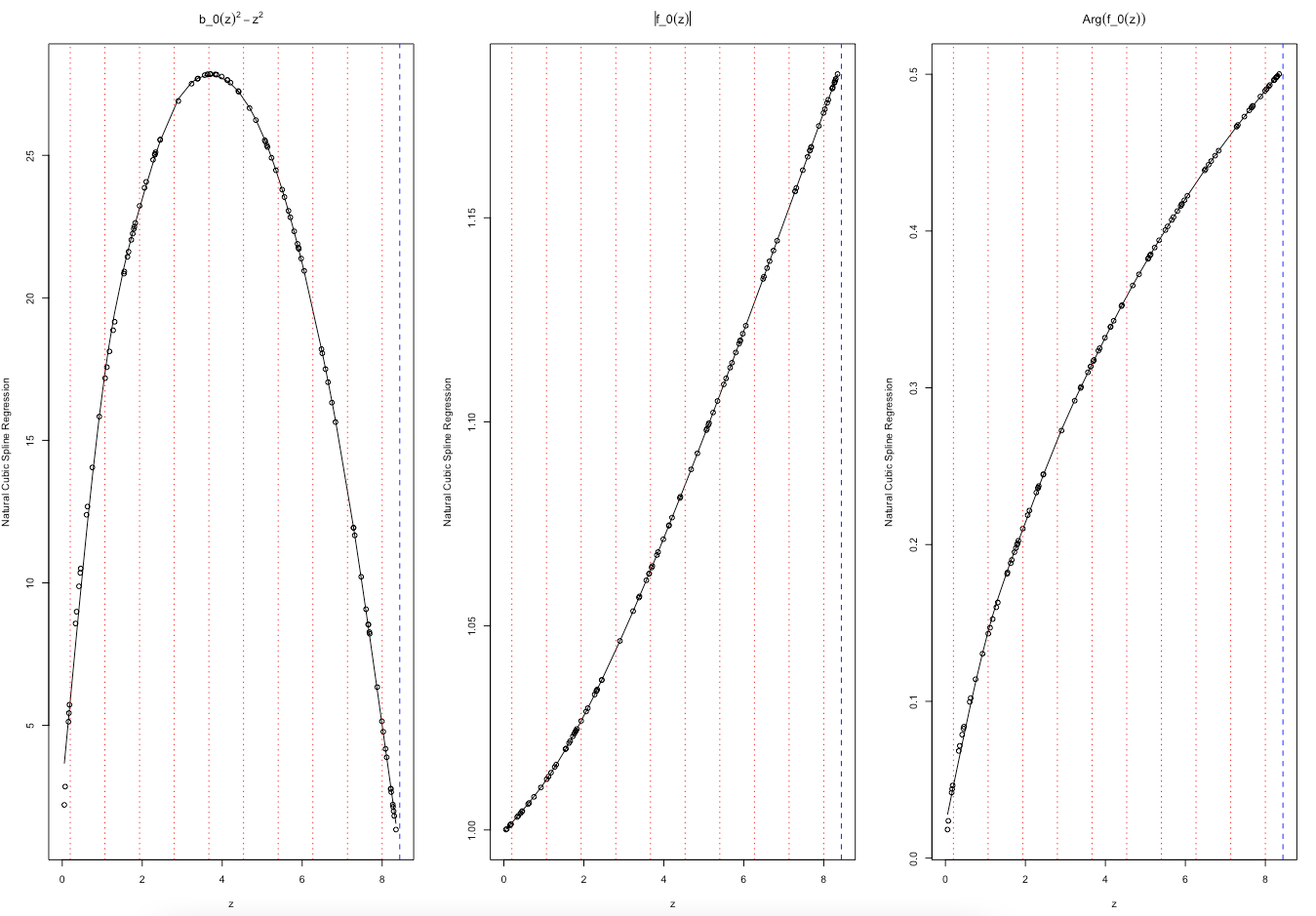}
\caption{\footnotesize{The estimates of critical collapse functions corresponding to $\gamma$-solution of hyperbolic case based on natural cubic spline
regression from a training sample of size $n=100$. The  red lines show the locations of the knots.}}
 \label{hy_ga_nat}
\end{figure}

\begin{figure}
\includegraphics[width=1.2\textwidth,center]{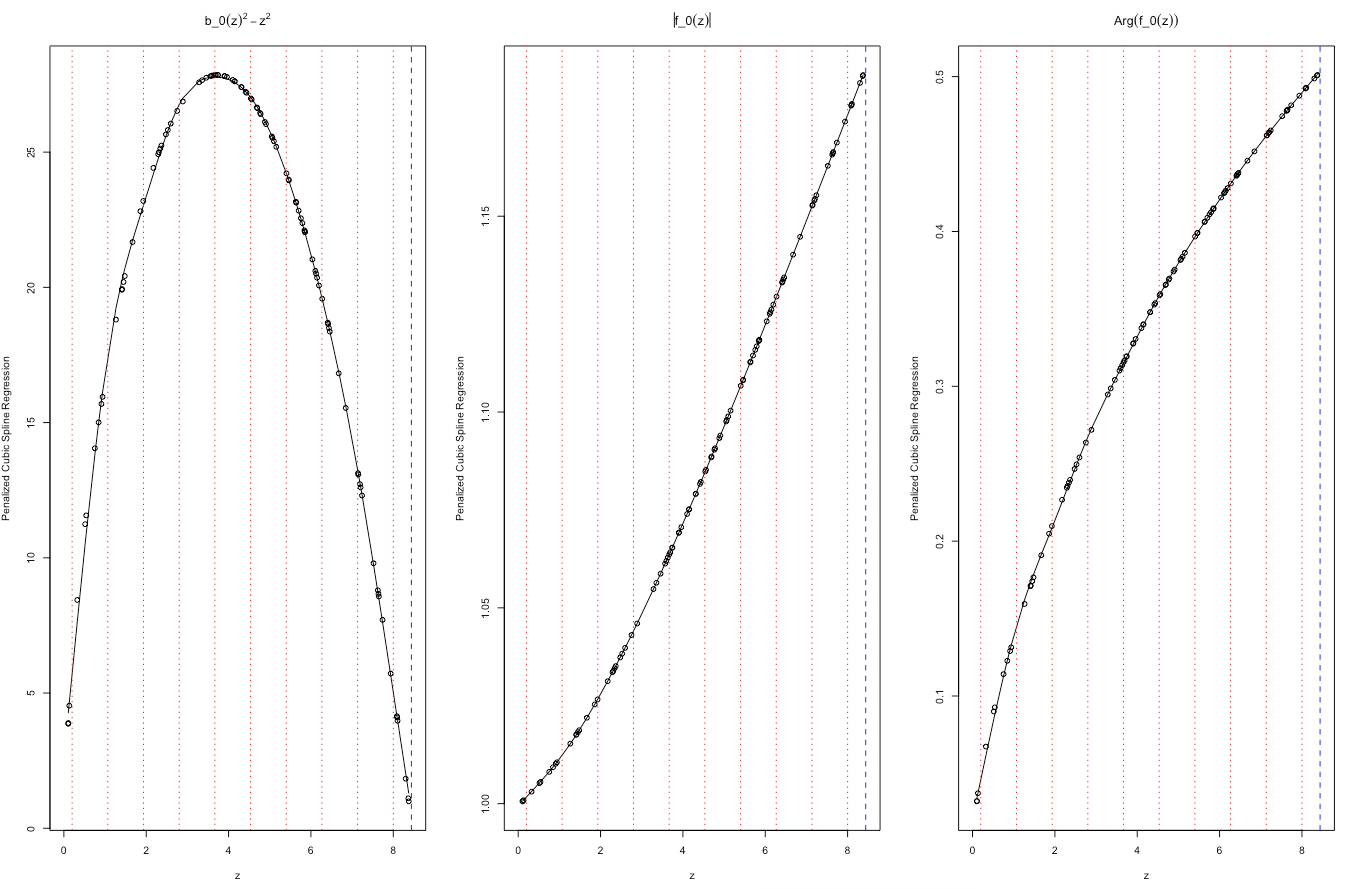}
\caption{\footnotesize{The estimates of critical collapse functions corresponding to $\gamma$-solution of hyperbolic case based on penalized B-spline 
regression from a training sample of size $n=100$. The  red lines show the locations of the knots.}}
 \label{hy_ga_pen}
\end{figure}

We show the performance of the proposed methods in estimating the critical collapse functions in both elliptic 
and hyperbolic cases in the following figures  and tables.
Figures \ref{hy_al_trunc}, \ref{el_trunc}, \ref{hy_be_trunc} and \ref{hy_ga_trunc} compare the estimation of critical collapse functions  using truncated power basis splines of orders $M=\{1,2,3\}$ in both elliptic ad hyperbolic spaces.
Owing to the advantages of the cubic splines, Tables \ref{absf_truncated}-\ref{argf_penalized}
show the $\sqrt{\text{MSE}}$ of truncated power basis spline, natural spline and penalized B-spline  of order $M=3$.
In each table, we illustrate the performance of the estimators using $K=\{3,5,8,10,15,20\}$ knots of equally 
spaced in the domain of each function.
From Tables \ref{absf_truncated}-\ref{argf_penalized}, we observe that the prediction error of almost all proposed spline models are  less than $10^{-2}$  such that
the proposed models can be considered unbiased estimators for the critical collapse functions in both elliptic and hyperbolic spaces 
(See also Figures \ref{hy_al_trunc}-\ref{hy_ga_pen}). 
We observe the $\sqrt{\text{MSE}}$ of the truncated power basis model is smaller than those of natural spline
 and penalized B-spline models.  
We also observe that penalized B-spline estimators almost always outperform the natural spline counterparts
 in predicting the elliptic and hyperbolic cases.  
Note that introducing more knots does not necessarily improve the performance of the truncated power basis model. There are several reasons for this fact. 
When $K$ increases, the multicollinearity problem arises in design matrix and consequently the estimation performance of the truncated power basis may not increase or even may decrease (e.g.,  Table \ref{absf_truncated} in the case of $\gamma$-solution).
Second, the higher polynomial orders may force the truncated power basis to change erratically  in the extreme regions while the critical function may not be highly nonlinear in the extreme regions. In this situation, we recommend the natural cubic spline smoother which imposes restrictions on the higher  orders in the extreme regions.
Given the training data, when $K$ increases,  no (or very few) training data  may be found in some regions. Thus, the estimation performance of the truncated power basis decreases in these regions. To handle this problem, we recommend the penalized B-spline smoother where sequentially selects the optimal number of knots in estimation procedure.
An advantage of the truncated power basis spline is that the method gives us a closed-form, 
continuous estimator with continuous derivatives of orders $M=\{1,2\}$ in predicting  the unperturbed 
critical collapse functions. This closed form and continuously differentiable estimators can play a high importance 
in paving the path for researchers to analytically find the critical solutions, critical exponents and mass of black holes. 

Given the positive part function  from \eqref{positive},  we define $u(z) = (z)^3_+$.  
The closed forms of critical collapse functions based on truncated power basis splines of order $M=3$ 
 in elliptic space with knots $\{l_1,\ldots,l_5\}=\{0.1,0.637,1.175,1.712,2.25\}$ are given by
\begin{align}\label{b_el_closed}\nonumber
\widehat{b_0(z)^2} =& 1 - 1.37 z + 31.02 z^2 - 91.22 z^3 + 88.17 u(z -l_1)  \\
&+ 3.47 u(z -l_2)  - 0.35 u(z -l_3) + 0.11 u(z -l_4) - 0.77 u(z -l_5),
\end{align}
\begin{align}\label{abdf_el_closed}\nonumber
\widehat{|f_0(z)|} =& 0.89 + 0.38 z -8.15 z^2 + 26.62 z^3 -26.32  u(z -l_1)  \\
&-0.45 u(z -l_2)  +0.15 u(z -l_3) -0.04 u(z -l_4) +0.22 u(z -l_5),
\end{align}
\begin{align}\label{argf_el_closed}\nonumber
\widehat{\arg f_0(z)} =& -0.0005 + 0.11 z - 2.13 z^2 + 8.32 z^3 -8.48  u(z -l_1)  \\
&+ 0.08 u(z -l_2)  +0.07 u(z -l_3) -0.01 u(z -l_4) +0.07 u(z -l_5).
\end{align}
The closed forms of critical collapse functions corresponding to $\alpha$-solution in hyperbolic space 
based on truncated power basis splines of order $M=3$ 
with knots $\{l_1,\ldots,l_5\}=\{0.2,0.475,0.75,1.025,1.3 \}$ are given by
\begin{align}\label{b_al_closed}\nonumber
\widehat{b_0(z)^2} =& 1 - 1.76 z -4.39 z^2 +2.96 z^3 + 6.02 u(z -l_1)  \\
&-6.53 u(z -l_2)  - 2.20 u(z -l_3) -1.10 u(z -l_4) +9.01 u(z -l_5),
\end{align}
\begin{align}\label{abdf_al_closed}\nonumber
\widehat{|f_0(z)|} =& 1.22 -0.0006 z -0.27 z^2 + 0.04 z^3 +0.18  u(z -l_1)  \\
&+0.13 u(z -l_2)  -0.33 u(z -l_3) -0.69 u(z -l_4) -0.95 u(z -l_5),
\end{align}
\begin{align}\label{argf_al_closed}\nonumber
\widehat{\arg f_0(z)} =& 0.61 -0.001 z +0.44 z^2 -0.003 z^3 +0.04  u(z -l_1)  \\
&-0.23 u(z -l_2)  -0.39 u(z -l_3) +0.22 u(z -l_4) +1.64 u(z -l_5).
\end{align}
The closed forms of critical collapse functions corresponding to $\beta$-solution in hyperbolic space 
based on truncated power basis splines of order $M=3$ 
with knots $\{l_1,\ldots,l_5\}=\{0.2,0.9,1.6,2.3,3 \}$ are given by
\begin{align}\label{b_be_closed}\nonumber
\widehat{b_0(z)^2} =& 1.05 - 2.20 z + 30.80 z^2 -50.54 z^3 + 49.82 u(z -l_1)  \\
&+1.12 u(z -l_2)  - 0.38 u(z -l_3) +0.11 u(z -l_4) -0.90 u(z -l_5),
\end{align}
\begin{align}\label{abdf_be_closed}\nonumber
\widehat{|f_0(z)|} =& 1 -0.04 z +0.15 z^2 + 0.05 z^3 -0.15  u(z -l_1)  \\
&+0.1 u(z -l_2)  -0.004 u(z -l_3) +0.005 u(z -l_4) -0.02 u(z -l_5),
\end{align}
\begin{align}\label{argf_be_closed}\nonumber
\widehat{\arg f_0(z)} =& 0.08 -0.28 z +4.46 z^2 -7.78 z^3 +7.81  u(z -l_1)  \\
&+0.02 u(z -l_2)  -0.05 u(z -l_3) +0.01 u(z -l_4) -0.11 u(z -l_5).
\end{align}
The closed forms of critical collapse functions corresponding to $\gamma$-solution in hyperbolic space 
based on truncated power basis splines of order $M=3$ 
with knots $\{l_1,\ldots,l_5\}=\{1,2.5,4,5.5,7\}$ are given by
\begin{align}\label{b_ga_closed}\nonumber
\widehat{b_0(z)^2} =& 1.01 +26.62 z -14.61 z^2 +4.52 z^3 -4.41  u(z -l_1)  \\
&-0.048 u(z -l_2)  - 0.047 u(z -l_3) +0.0049 u(z -l_4) -0.02 u(z -l_5),
\end{align}
\begin{align}\label{abdf_ga_closed}\nonumber
\widehat{|f_0(z)|} =& 0.99 -.04 z +.01 z^2 + .0007 z^3 +.0006  u(z -l_1)  \\
&-.0009 u(z -l_2)  +.0001 u(z -l_3)+.00003 u(z -l_4) -.00001 u(z -l_5).
\end{align}
\begin{align}\label{argf_ga_closed}\nonumber
\widehat{\arg f_0(z)} =& .009 +.203 z -.108 z^2 -.0334 z^3 -.0327  u(z -l_1)  \\
&-.00005 u(z -l_2)  -.00044 u(z -l_3) +.00006 u(z -l_4) -.000198 u(z -l_5).
\end{align}

\section{Summary and Concluding Remarks}\label{sec:con}

In this paper, we investigated the properties of the spline regression models in estimating the nonlinear critical collapse functions for axion-dilaton system in elliptic and hyperbolic cases in four dimensions. 
The spline regression models include truncated power basis, natural cubic spline and penalized B-splines. 
The truncated power basis and natural spline use knots over the domain of a nonlinear function to accommodate  the local changes in the prediction using the basis functions and hence we do not require to adjust the global pattern of the models. 
In addition to using knots, the penalized B-spline regression model uses the constrained optimization  to reach the optimal level of smoothness and avoid overfitting problem.  Through various numerical studies, we evaluated the performance of the proposed spline models in estimating the critical collapse functions. We observed that the error of the developed estimators, on average, are almost less than $10^{-2}$ such that all the estimators can be considered as an unbiased estimators for the critical collapse functions. 
In addition to the accuracy of the proposed models, the truncated power basis spline provides us a closed-form, 
continuous estimator with continuous derivatives of orders $M=\{1,2\}$ in predicting  the unperturbed 
critical collapse functions. This analytical estimators can be of high importance to pave the path for finding analytically the solutions, critical exponents and mass of black holes. 

\section*{Acknowledgment}
The authors would like to thank the anonymous referee and the editor for their constructive comments which improved the quality and presentation of the manuscript. 
Ehsan Hatefi would like to thank L. Alvarez-Gaume, R. Antonelli, E. Hirschmann, A. Sagnotti and R. J. López-Sastre for their valuable comments and insights. Ehsan Hatefi acknowledges the research support provided under International grant of Maria Zambrano. Armin Hatefi acknowledges the research support of the Natural Sciences and Engineering Research Council of Canada (NSERC).
 

\newpage
\section*{Appendix}



\vspace{1cm}

\begin{table}[ht]
\centering
\footnotesize{\begin{tabular}{lccccc}
  \hline
  $K$   &   Elliptic &            & Hyperbolic &  \\ 
  \cline{3-5} 
        &            & $\alpha$-solution   & $\beta$-solution    & $\gamma$-solution \\ 
\hline
 3 & 0.0222907 & 0.0004868 & 0.0332154 & 0.0679797 \\ 
 5 & 0.0049895 & 0.0000634 & 0.0135313 & 0.0680210 \\ 
 8 & 0.0005913 & 0.0000074 & 0.0031689 & 0.0664356 \\ 
 10 & 0.0001874 & 0.0000034 & 0.0016043 & 0.0649515 \\ 
 15 & 0.0004048 & 0.0000037 & 0.0004153 & 0.0614526 \\ 
 20 & 0.0001411 & 0.0000048 & 0.0001062 & 0.0596425 \\ 
   \hline
\end{tabular}
\caption{\footnotesize{ The $\sqrt{\text{MSE}}$ of truncated power basis method of orders $M=3$ with $K=\{3,5,\ldots,20\}$ knots in estimating 
the critical collapse response function $g(z)=b_0^2(z)-z^2$ in elliptic and hyperbolic cases based on a training sample of size $n=100$.}}
\label{b_truncated}}
\end{table}

%
%
%
%

\vspace{1cm}

\begin{table}[ht]
\centering
\footnotesize{\begin{tabular}{lccccc}
  \hline
  $K$   &   Elliptic &            & Hyperbolic &  \\ 
  \cline{3-5} 
        &            & $\alpha$-solution   & $\beta$-solution    & $\gamma$-solution \\ 
\hline
 3 & 0.0007897 & 0.0009433 & 0.0015890 & 0.0006771 \\ 
 5 & 0.0003643 & 0.0001183 & 0.0015586 & 0.0006875 \\ 
 8 & 0.0001263 & 0.0000113 & 0.0005004 & 0.0006727 \\ 
 10 & 0.0000584 & 0.0000089 & 0.0002765 & 0.0006613 \\ 
 15 & 0.0000931 & 0.0000089 & 0.0000794 & 0.0006356 \\ 
 20 & 0.0000290 & 0.0000099 & 0.0000194 & 0.0006224 \\ 
   \hline
\end{tabular}
\caption{\footnotesize{ The $\sqrt{\text{MSE}}$ of truncated power basis method of orders $M=3$ with $K=\{3,5,\ldots,20\}$ knots in estimating 
the critical collapse response function $g(z)=\arg f_0(z)$ in elliptic and hyperbolic cases based on a training sample of size $n=100$.}}
\label{argf_truncated}}
\end{table}

\vspace{1cm}

\begin{table}[ht]
\centering
\footnotesize{\begin{tabular}{lccccc}
  \hline
  $K$   &   Elliptic &            & Hyperbolic &  \\ 
  \cline{3-5} 
        &            & $\alpha$-solution   & $\beta$-solution    & $\gamma$-solution \\ 
\hline
 3 & 0.0583561 & 0.0100865 & 0.1012991 & 1.4876654 \\ 
 5 & 0.0452239 & 0.0045656 & 0.0472993 & 0.8784532 \\ 
 8 & 0.0274994 & 0.0025943 & 0.0439414 & 0.6357770 \\ 
 10 & 0.0209406 & 0.0020206 & 0.0396018 & 0.5490553 \\ 
 15 & 0.0128638 & 0.0014594 & 0.0323031 & 0.4459167 \\ 
 20 & 0.0103324 & 0.0012106 & 0.0274119 & 0.3893207 \\ 
   \hline
\end{tabular}
\caption{\footnotesize{ The $\sqrt{\text{MSE}}$ of cubic natural spline method  with $K=\{3,5,\ldots,20\}$ knots in estimating 
the critical collapse response function $g(z)=b_0^2(z)-z^2$ in elliptic and hyperbolic cases based on a training sample of size $n=100$.}}
\label{b_natural}}
\end{table}

%

\newpage
\vspace{1cm}

\begin{table}[ht]
\centering
\footnotesize{\begin{tabular}{lccccc}
  \hline
  $K$   &   Elliptic &            & Hyperbolic &  \\ 
  \cline{3-5} 
        &            & $\alpha$-solution   & $\beta$-solution    & $\gamma$-solution \\ 
\hline
 3 & 0.0145808 & 0.0182347 & 0.0175141 & 0.0077258 \\ 
 5 & 0.0035752 & 0.0040297 & 0.0046769 & 0.0045959 \\ 
 8 & 0.0007587 & 0.0018753 & 0.0047694 & 0.0033778 \\ 
 10 & 0.0002901 & 0.0014674 & 0.0047028 & 0.0029812 \\ 
 15 & 0.0001175 & 0.0010645 & 0.0041953 & 0.0025177 \\ 
 20 & 0.0001081 & 0.0008880 & 0.0037143 & 0.0022268 \\ 
   \hline
\end{tabular}
\caption{\footnotesize{ The $\sqrt{\text{MSE}}$ of cubic natural spline method with $K=\{3,5,\ldots,20\}$ knots in estimating 
the critical collapse response function $g(z)=\arg f_0(z)$ in elliptic and hyperbolic cases based on a training sample of size $n=100$.}}
\label{argf_natural}}
\end{table}

\vspace{1cm}

\begin{table}[ht]
\centering
\footnotesize{\begin{tabular}{lccccc}
  \hline
  $K$   &   Elliptic &            & Hyperbolic &  \\ 
  \cline{3-5} 
        &            & $\alpha$-solution   & $\beta$-solution    & $\gamma$-solution \\ 
\hline
 3 & 0.0594606 & 0.0088410 & 0.1029377 & 1.6690824 \\ 
 5 & 0.0378520 & 0.0029216 & 0.0339346 & 0.9070288 \\ 
 8 & 0.0188786 & 0.0013406 & 0.0275224 & 0.6636289 \\ 
 10 & 0.0132785 & 0.0009962 & 0.0256748 & 0.5671875 \\ 
 15 & 0.0093517 & 0.0005840 & 0.0197771 & 0.4318121 \\ 
 20 & 0.0069107 & 0.0004166 & 0.0166023 & 0.3937617 \\ 
   \hline
\end{tabular}
\caption{\footnotesize{ The $\sqrt{\text{MSE}}$ of penalized cubic B-spline method  with $K=\{3,5,\ldots,20\}$ knots in estimating 
the critical collapse response function $g(z)=b_0^2(z)-z^2$ in elliptic and hyperbolic cases based on a training sample of size $n=100$.}}
\label{b_penalized}}
\end{table}

%

\vspace{1cm}

\begin{table}[ht]
\centering
\footnotesize{\begin{tabular}{lccccc}
  \hline
  $K$   &   Elliptic &            & Hyperbolic &  \\ 
  \cline{3-5} 
        &            & $\alpha$-solution   & $\beta$-solution    & $\gamma$-solution \\ 
\hline
 3 & 0.0123415 & 0.0188509 & 0.0178855 & 0.0095065 \\ 
 5 & 0.0030677 & 0.0025941 & 0.0035336 & 0.0051307 \\ 
 8 & 0.0006357 & 0.0009867 & 0.0028396 & 0.0038707 \\ 
 10 & 0.0002447 & 0.0007396 & 0.0029189 & 0.0033959 \\ 
 15 & 0.0000877 & 0.0004281 & 0.0025423 & 0.0027116 \\ 
 20 & 0.0000719 & 0.0003076 & 0.0022254 & 0.0025339 \\ 
   \hline
\end{tabular}
\caption{\footnotesize{ The $\sqrt{\text{MSE}}$ of penalized cubic B-spline method with $K=\{3,5,\ldots,20\}$ knots in estimating 
the critical collapse response function $g(z)=\arg f_0(z)$ in elliptic and hyperbolic cases based on a training sample of size $n=100$.}}
\label{argf_penalized}}
\end{table}


\newpage
\begin{figure}
\includegraphics[width=1.2\textwidth,center]{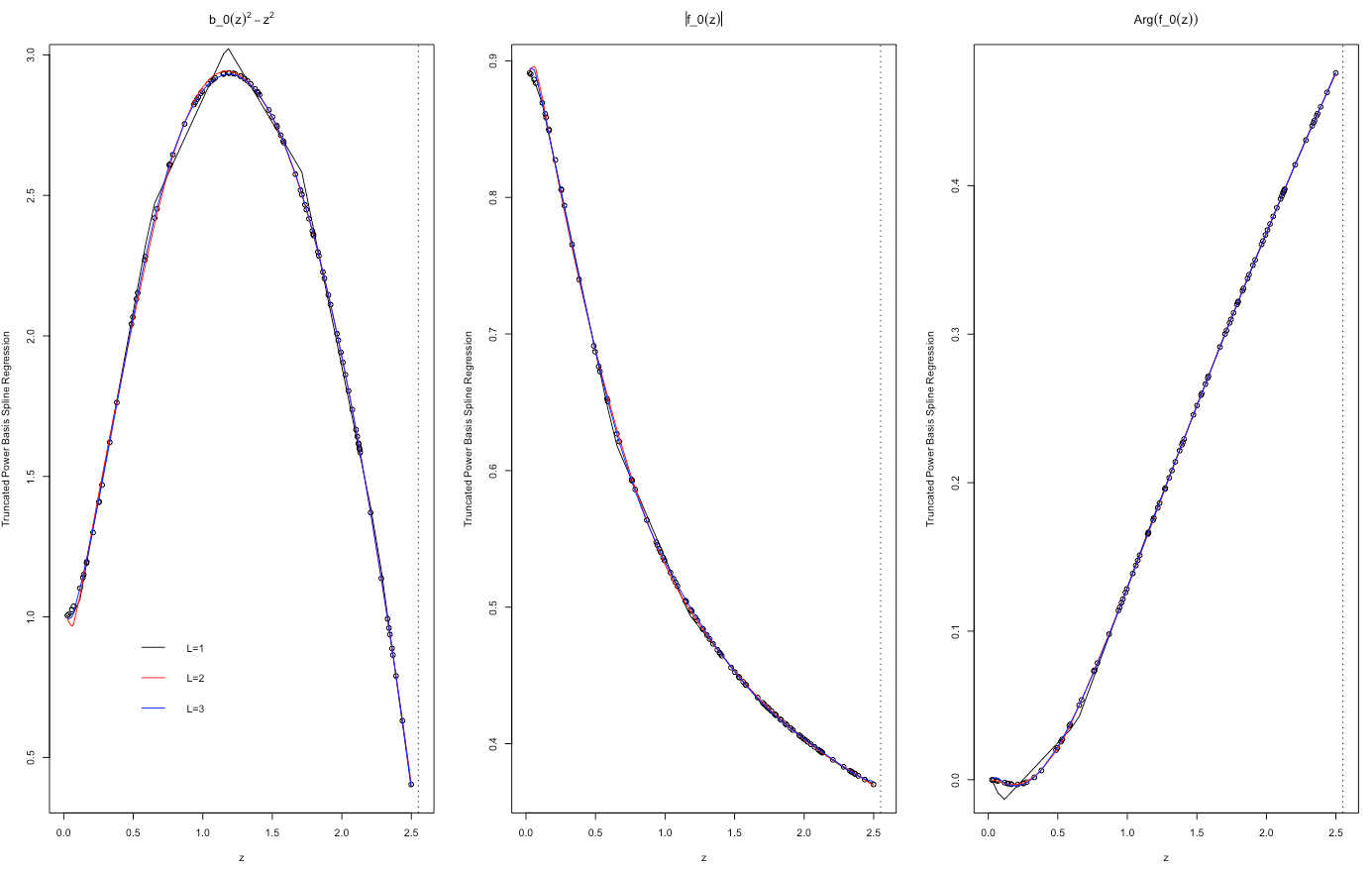}
\caption{\footnotesize{The estimates of critical collapse functions based on truncated power basis 
regression of orders $m=\{1,2,3\}$ in elliptic case from a training sample of size $n=100$.}}
 \label{el_trunc}
\end{figure}

\newpage

\newpage

%
%
%

\newpage

\newpage

\newpage

\newpage

\newpage

\newpage


\end{document}